\documentclass{IEEEtran}
\usepackage[utf8]{inputenc} 
\usepackage[T1]{fontenc}
\usepackage{url}
\usepackage{ifthen}
\usepackage{cite}
\usepackage{amssymb}
\usepackage{amsfonts}
\usepackage{amsthm}
\usepackage[cmex10]{amsmath}
\usepackage{dsfont}
\usepackage{balance}
\usepackage{xfp}
\usepackage{makecell}

\usepackage{graphicx}
\usepackage{hyperref}                                   
\hypersetup{colorlinks=true,allcolors=blue}
\usepackage{xargs}                                      
\usepackage{xcolor}                                     
\usepackage{algorithm}                                  
\usepackage{algpseudocode}                              
\usepackage[printonlyused]{acronym}                     
\usepackage{enumitem}                                   
\usepackage{lipsum}                                     
\usepackage{xspace}
\newtheorem{theorem}{Theorem}
\newtheorem{corollary}{Corollary}

\newtheorem{lemma}{Lemma}

\newtheorem{proposition}{Proposition}
\newtheorem{claim}{Claim}

\acrodef{HT}[HT]{Hypothesis Testing}
\acrodef{DHT}[DHT]{Distributed Hypothesis Testing}
\acrodef{SHT}[SHT]{Sequential \ac{HT}}
\acrodef{KLD}[KLD]{Kullback-Leibler Divergence}
\acrodef{DM}[DM]{Decision-Maker}
\acrodef{DMC}[DMC]{Discrete Memoryless Channel}
\acrodef{PDF}[PDF]{Probability Density Function}
\acrodef{PMF}[PMF]{Probability Mass Function}
\acrodef{CDF}[CDF]{Cumulative Distribution Function}
\acrodef{LLR}[LLR]{Log-Likelihood Ratio}
\acrodef{LLRT}[LLRT]{\ac{LLR} Test}
\acrodef{AI}[AI]{Artificial Intelligence}
\acrodef{ML}[ML]{Machine Learning}
\acrodef{ABR}[ABR]{Average Bayes Risk}
\acrodef{AEP}[AEP]{Asymptotic Equipartition Property}
\acrodef{MWDT}[MWDT]{Minimal Weight Decision Tree}
\acrodef{TVD}[TVD]{Total Variation Distance}
\acrodef{MGF}[MGF]{Moment-Generating Function}
\acrodef{ISIT}[ISIT]{IEEE International Symposium on Information Theory}
\acrodef{NHSHT}[NHSHT]{Non-Homogeneous \ac{SHT}}
\acrodef{CASHT}[CASHT]{\ac{CA} \ac{SHT}}
\acrodef{SPRT}[SPRT]{Sequential Probability Ratio Test}
\acrodef{MSPRT}[MSPRT]{Multihypothesis \ac{SPRT}}
\acrodef{CA}[CA]{Cost-Aware}
\acrodef{BPB}[BPB]{Bit-Per-Buck}
\acrodef{MEF}[MEF]{Mean Excess Function}
\acrodef{LP}[LP]{Linear Programming}
\acrodef{RL}[RL]{Reinforcement Learning}
\acrodef{POMDP}[POMDP]{Partially Observable Markov Decision Process}

\acrodef{PHI}[PHI]{Pruning Hypotheses Iteratively}
\acrodef{DELTA}[DELTA]{Distribution-based Early Labeling for Tapered Acquisitions}
\acrodef{IOTA}[IOTA]{Independent One-at-a-Time Action}
\acrodef{SEF}[SEF]{Scalar Exponential Family}
\acrodef{MGF}[MGF]{Moment-Generating Function}

\newcommand{\CACH}{\ac{CA}-Chernoff\xspace}
\newcommand{\CANJ}{\ac{CA}-NJ1\xspace}

\def\HT{\ac{HT}\xspace}

\def\SHT{\ac{SHT}\xspace}
\def\KLDtxt{\ac{KLD}\xspace}
\def\KLDstxt{\acp{KLD}\xspace}
\def\DM{\ac{DM}\xspace}

\def\PDF{\ac{PDF}\xspace}

\def\CDF{\ac{CDF}\xspace}
\def\CDFs{\acp{CDF}\xspace}

\def\LLR{\ac{LLR}\xspace}
\def\LLRs{\acp{LLR}\xspace}

\def\SPRT{\ac{SPRT}\xspace}

\def\ABR{\ac{ABR}\xspace}

\def\TVDtxt{\ac{TVD}\xspace}

\def\CASHT{\ac{CASHT}\xspace}
\def\MSPRT{\ac{MSPRT}\xspace}

\def\CA{\ac{CA}\xspace}

\def\LP{\ac{LP}\xspace}

\newif\ifBlind
\Blindtrue
\Blindfalse 

\newif\ifShowProofSkech
\ShowProofSkechtrue

\newif\ifPaperAward
\PaperAwardtrue
\PaperAwardfalse 

\newif\ifSubmittedForReview
\SubmittedForReviewtrue
\SubmittedForReviewfalse 

\newif\ifShowSupp
\ShowSupptrue

\newcommandx{\myVec}[2][2=]{{\underline{\smash{#1}}_{#2}}}                      
\newcommandx{\myMat}[2][2=]{{\mathbf{#1}_{#2}}}                                 
\newcommand{\component}[2]{[#1]_{#2}}
\newcommandx{\vectorComponent}[3][3=]{\component{\myVec{#1}[#3]}{#2}}           
\newcommandx{\matrixComponent}[4][4=]{\component{\myMat{#1}[#4]}{#2,#3}}        
\newcommand{\KLD}[2]{\mathcal{D}_{KL} ( #1 \Vert #2 )}                          

\newcommand{\expVal}[1]{\mathbb{E}\left[#1\right]}                              
\newcommand{\expValDist}[2]{\mathbb{E}_{#2}\left[#1\right]}
\newcommand{\indicator}[1]{ \mathds{1}{\left\{#1\right\}} }                     
\newcommand{\prob}[1]{\operatorname{\mathbb{P}}\left( #1 \right)}               
\newcommand{\bigO}[1]{\operatorname{\mathcal{O}}\left( #1 \right)}              
\newcommand{\littleO}[1]{\operatorname{o}\left( #1 \right)}                     
\newcommand{\bigTheta}[1]{\operatorname{\Theta}\left( #1 \right)}               
\newcommand{\hII}{H_i}

\newcommand{\argmax}[2]{\mathop{\operatorname{argmax}}_{#2} #1}

\newcommand{\abs}[1]{ | #1 |}

\newcommandx{\norm}[2][2=2]{\| #1 \|_{#2}}
\newcommand{\indexedSet}[3]{\myVec{#1}_{#2}^{#3}}
\newcommand{\TVD}[1]{ \| #1 \|_{\mathrm{TV}} }
\newcommand{\TVDij}[0]{\TVD{f_i^a-f_j^a}}

\newcommand{\probSimplex}[1]{\operatorname{\mathcal{PS}}(#1)}
\newcommand{\probSimplexA}[0]{\probSimplex{\abs{\mathcal{A}}}}
\newcommand{\incompleteBetaFunc}[3]{\mathrm{B}\left(#1; #2, #3\right)}
\newcommand{\partialDerive}[1]{\frac{\partial}{\partial#1}}
\newcommand{\partialDeriveSecond}[1]{\frac{\partial^2}{\partial#1^2}}
\def\abstain{\mathrm{\textbf{abstain}}}

\date{}

\title{On Cost-Aware Designs for Sequential Hypothesis Testing}
\ifBlind
    \author{%
      \IEEEauthorblockN{Anonymous Authors}
      \IEEEauthorblockA{%
        }
    }
\else
    \author{
        \IEEEauthorblockN{George Vershinin, Asaf Cohen, and Omer Gurewitz}
        
        \IEEEauthorblockA{The School of Electrical and Computer Engineering,
                        Ben-Gurion University of the Negev, Israel
                        \newline
                        georgeve@post.bgu.ac.il, \{coasaf, gurewitz\}@bgu.ac.il}
        \thanks{Parts of this work were presented at the IEEE International Conference on Acoustics, Speech, and Signal Processing, ICASSP 2026, \cite{vershinin2025activesequentialhypothesistesting}}
    }
\fi

\ifSubmittedForReview
\fi

\begin{document}

\maketitle
\begin{abstract}
We introduce Cost-Aware (CA) Sequential Hypothesis Testing (CASHT), in which an active decision-maker selects sensing actions with different, random costs to identify the true hypothesis under an average-error constraint $\delta$, while minimizing the expected total cost (rather than the number of samples). For fixed costs, we prove that the optimal expected total cost scales as $\Theta(\log(1/\delta))$, and is achievable by Multihypothesis Sequential Probability Ratio Test-based procedures. We show that the CA design principle is to maximize the ratio of expected information gain to expected cost under the policy-induced action distribution. Guided by this principle, we adapt two classic policies to the CA setting and establish their asymptotic optimality. We then treat random costs under two revelation models: ex-post, where costs are disclosed only after a sample is obtained, and the cost-error tradeoff coincides with the fixed-cost case, and ex-ante, where costs accrue before acquisition, and the decision maker may cancel an action mid-operation. For the ex-ante model, we characterize when cancellation lowers the total cost and analyze several cost distributions in detail. Simulations confirm our findings that the CA variants consistently reduce total cost relative to their classic counterparts, and when action cancellation helps or hurts.

\end{abstract}

\section{Introduction}
\label{section: introduction}
\HT is a central tool in statistical inference in which a \DM determines whether observed data provides sufficient evidence to decide among competing hypotheses.
This technique is widely used across highly diverse domains, such as physics, manufacturing, finance, environmental science, and medicine, and its relevance has only grown with the rise of artificial intelligence that must make reliable, data-driven decisions at scale.

Classic \HT operates by selecting which of two distributions best explains a fixed-length batch of data collected in advance.
In many modern computing systems, such as anomaly detectors and networked sensors, the \DM continuously collects data in real time to detect events or classify states, which naturally led to the creation of the \SHT.

Wald’s pioneering work in \cite{Wald_1945_SHT} formalized the online binary decision problem and introduced the \SPRT.
In the \SPRT, the \DM continuously accumulates \LLRs while they remain between two predefined decision thresholds, terminating once either threshold is crossed.
In these settings, the \SPRT is optimal in the sense that, for any prescribed error probabilities, it minimizes the expected number of samples among sequential tests satisfying the same error constraints \cite{Wald_Wolfowitz_1948_SPRT_Optimality}, a quantity that is routinely interpreted, \emph{and even named}, as detection delay.
This notion of optimality inspired the \MSPRT \cite{Armitage1950_SHT_MultipleHypotheses, Dragalin_etAl_1999_MSPRT_AsympOpt, Dragalin_etAl_2000_MSPRT_MeanSamplesApprox}, which extends the \SPRT to multiple hypotheses while retaining the same sample scaling laws.

In many scenarios, the \DM has numerous information sources to query, allowing it to choose which source to probe next (e.g., which router to monitor for cyber-attacks or which diagnostic test to run next), thereby enabling active sensing.
Accordingly, the \MSPRT has also been adopted for active settings to identify the correct hypothesis, with actions guided to use as few samples as possible, e.g., \cite{Chernoff1959SequentialHT, Naghshvar_Javidi2013_SHT_DynamicProgramming, Nitinawarat2013_SHT_Argmax2_KLD_wProofs, Cohen_Zhao2015_SHT_AnomalyDetection, Bai_Katewa_Gupta_Huang2015_Stochastic_Source_Selection, Citron_Cohen_Zhao2024_DGF_on_Hidden_Markov_Chains}.
Other extensions leveraged the \MSPRT itself as a mechanism to iteratively eliminate hypotheses that are inconsistent with observed data, e.g., \cite{Gan_Jia_Li2021_Decision_Tree_SHT, vershinin2025multistageactivesequentialhypothesis, vershinin2025iterativehypothesispruningdistributionbased}, again minimizing sample count.
Even recent learning-based centralized and distributed methods focus on the number of samples (or a function of it), e.g., \cite{Gurevich2019_EEST, Szostak2024_DeepLearining2, stamatelis2024_DeepLearining3}.

This shared focus on the expected number of samples encodes the assumption that every action costs the same.
However, in practical settings that motivate active testing, this assumption breaks.
A physician weighs a cheap blood test against an expensive biopsy; a monitor weighs a local read against a remote query; a generated sample is billed by the number of tokens it consumes.
Namely, costs such as latency, energy, and billing vary across actions and may differ by orders of magnitude.
Additionally, the action cost may not reflect how informative the action is or its corresponding sample.
When action costs differ, minimizing the number of samples is no longer the same as minimizing what the \DM actually pays, and the two objectives can be arbitrarily far apart.

\SHT with acquisition costs has been approached from a learning perspective.
Recognizing that action-dependent costs render this problem a Partially Observable Markov Decision Process, \cite{Joseph_DeepLearining1} empirically showed that a policy learned via reinforcement learning can outperform the seminal Chernoff scheme in \cite{Chernoff1959SequentialHT} in terms of total cost.
This gain, however, is demonstrated only numerically rather than analytically, and the question of how close any policy, learned or otherwise, comes to the best achievable cost remains open, as the achievable cost has never been characterized.

In this paper, we formulate the \CASHT problem.
Each action incurs a positive random acquisition cost drawn afresh from a known distribution, and the \DM minimizes the expected total cost subject to an average-error constraint.
This formulation extends the classic \SHT objective by measuring performance directly in terms of the cost actually incurred by the \DM.
Specifically, when all actions incur the same cost, minimizing the expected total cost is equivalent to minimizing the expected number of samples.
Moreover, when the acquisition cost represents the time required to obtain a sample, the expected total cost coincides with the expected \emph{literal} detection delay.

Optimizing the expected total cost is subtler than it first appears.
The prominent strategy in \SHT, pioneered by Chernoff in \cite{Chernoff1959SequentialHT}, focuses on taking actions that maximize the average information gain per sample.
A na\"ive \CA analogue for this strategy is to re-score actions by their information gain per unit cost.
As we show later, however, this seemingly natural per-action score re-scaling does not, in general, align with the \CASHT goal.

In this work, we consider two cost revelation models.
In the first model, the ex-post model, costs are revealed only after the sample is acquired, and the \DM commits to an action without knowing the exact cost it will incur.
The second is the ex-ante model, where the cost accrues over time and can be observed as it does.
In some scenarios, e.g., when costs represent delay, the ex-ante model allows the \DM to react to unfavorable cost realizations by canceling ongoing actions, paying only a partial cost, and receiving no sample, and then continuing to operate as usual, either by selecting a new action or retrying the aborted one.
Note that without this preemption ability, the two cost-revelation models coincide in their analysis.

Our contributions are as follows:
\begin{itemize}
    \item \textbf{Asymptotic-optimal cost, and the design principle it dictates.}
    We prove matching asymptotic bounds on the optimal expected total cost under an error constraint $\delta$, resulting in an oracle bound attainable by \MSPRT-based procedures (Theorem \ref{theorem: expected total cost bounds}). 
    The leading constant of these bounds dictates the \CA policy design principle of maximizing the ratio of expected information gain to expected cost (Corollary \ref{corollary: expected information gain per expected cost optimization goal}).
    
    \item \textbf{Two \CA adaptive stochastic procedures.}
    We leverage the design principle in two asymptotically-optimal procedures: \CACH, that extends the Chernoff scheme (Theorem \ref{theorem: Chernoff order optimality}), and \CANJ, based on Policy 1 of \cite{Naghshvar_Javidi2013_SHT_DynamicProgramming}, which precedes the \CACH scheme with an exploration phase to improve stability (Theorem \ref{theorem: NJ1 order optimality}).
    Both attain the optimal scaling; in simulation, their costs are close, with \CANJ incurring a small exploration penalty.

    \item \textbf{Preemption Analysis.}
    We analyze the effect of action cancellation in the ex-ante model on the expected total cost for costs drawn afresh at each acquisition.
    We derive a cost-overshoot criterion characterizing when cancellation is beneficial (Lemma \ref{lemma: kappa leq expValC}), and analyze Erlang, Hyperexponential, Pareto, and Log-Logistic costs as canonical examples (Propositions \ref{proposition: Erlang Example}-\ref{proposition: LogLogistic Example}).
\end{itemize}
\section{System Model}
\label{section: system model}

\subsection{Notation}
\label{subsection: notation}
All vectors in this manuscript are underlined (e.g., $\myVec{x}$).
For the all-zeroes vector, we write $\myVec{0}$, and, similarly, $\myVec{1}$ is the all-ones vector.
Vector components are specified in conjunction with square brackets, e.g., $\vectorComponent{y}{i}$ is $\myVec{y} $’s $i^{\mathrm{th}}$ component.
We write $\myVec{x}\leq \myVec{y}$ as a shorthand notation for $\vectorComponent{x}{i}\leq \vectorComponent{y}{i}$ for all $i$.
We use $\indexedSet{x}{i}{j}$ as a shorthand notation for $(x_i, x_{i+1}, \dots, x_j)^T$, where $T$ is the transpose operation.

The expectation with respect to some random variable $X$ is denoted as $\expValDist{\cdot}{X}$.
For notational convenience, if clear from context, we will omit the subscript in the expectation, e.g., $\expVal{X}$ instead of $\expValDist{X}{X}$.
The \KLDtxt between two densities $f$ and $g$ is denoted as $\KLD{f}{g}$.
Unless explicitly specified (e.g., $\ln$), all logarithms in this manuscript are in base two.

As accepted in the literature, e.g., \cite[Chapter~11]{CoverThomas2006}, discrete distributions over $n$-sized alphabets will be regarded as vectors over the probability simplex, $\probSimplex{n} \triangleq \{ \myVec{x}\in\mathbb{R}^n: \myVec{x}^T\myVec{1} = 1, \myVec{x}\geq \myVec{0} \}$.

\subsection{Model}
\label{subsection: model}
We consider a phenomenon that cannot be observed directly, but whose manifestations can be observed.
The phenomenon is characterized by one of a finite set of hypotheses, $\mathcal{H} = \{0, 1, \ldots, H-1\}$.
A \DM can perform actions from a finite set $\mathcal{A} = \{1,2,\ldots,\abs{\mathcal{A}}\}$, such as issuing queries or conducting tests.
Upon selecting an action, the \DM observes a sample whose distribution depends on both the underlying hypothesis and the chosen action.
Consequently, different actions may induce different observation distributions for the same hypothesis.
Specifically, each hypothesis-action pair $(h, a)\in\mathcal{H}\times\mathcal{A}$ induces a probability distribution over the observation space denoted by $f_h^a(x) = f(x|h,a)$.
The distributions $\{f_h^a\}_{h, a}$ are assumed to be known by the \DM.
For notational simplicity, we focus on scalar samples; extending to non-scalar samples is straightforward.

If action $a\in\mathcal{A}$ is taken, a random cost $C_a\sim f_{C_a}$, whose support and \CDF are $[0, \infty)$ and $F_{C_a}$, respectively, is incurred.
We emphasize that this cost reflects the real-world characteristics of the sample taken (e.g., time to acquire the sample, billing, energy use), rather than its statistical or dynamic structure (e.g., informativeness, hypothesis-separation capability, moments).
The cost is drawn independently each time the action is applied, and is independent of the underlying unobservable phenomenon and of previous or future actions.
Furthermore, the cost densities $\{f_{C_a}\}_{a}$ are assumed to be known.

We consider two cost revelation models.
In the \textit{ex-post} model, cost realization is revealed only after an action is completed, so the \DM commits without knowing what it will pay.
In the \textit{ex-ante} model, the cost realization is revealed before the sample is acquired.
The ex-ante model permits a response that the ex-post model cannot: preemption.
In the non-preemptive case, the \DM commits to each action and receives both its sample and its full realized cost.
In the preemptive case, the \DM may abort an action in progress; the aborted action then yields the null symbol $\{\abstain\}$ in place of a sample, while still incurring the a function of the cost accrued up to the abort.
Specifically, motivated by delay, we assume that the cost of a preempted action equals the cost accrued until it is aborted.

\begin{figure}[!htbp]
    \centering
    \includegraphics[]{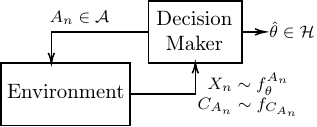}
    \vspace{-8pt}
    \caption[CASHT Model]{
        System model.
        The \DM seeks to identify the correct hypothesis indexed by $\theta\in\mathcal{H}$.
        By taking action $A_n$ at time step $n$, the \DM obtains a realization of $X_n\sim f_\theta^{A_n}$, and a random cost $C_{A_n}\sim f_{C_{A_n}}$ is incurred.
        }
    \label{fig: model}
\end{figure}

We assume no prior knowledge of the underlying hypothesis.
Namely, the prior probability of hypothesis $i$ (denoted by $\hII$) is $\prob{\theta = i} = \frac{1}{H}, \forall i \in \mathcal{H}$.
We further assume that the samples are independent and identically distributed conditioned on the action taken, with the distribution parameters depending on the true underlying hypothesis. 
The decision made is given by $\hat{\theta}\in\mathcal{H}$, i.e., $\hat{\theta} = i$ implies that $\hII$ is declared as true.
Figure \ref{fig: model} visualizes the model.

We make additional assumptions:
\begin{enumerate}[label=(A\arabic*)]
    \item (Separation) For any action $a\in \mathcal{A}$, there are at least two hypotheses $i$, $j\in\mathcal{H}$ with $\TVDij > 0$.
    \label{assumption: separation}
    \item (Validity) For all $i, j\in\mathcal{H}$, there is at least one action $a\in\mathcal{A}$ with $\TVDij > 0$.
    \label{assumption: validity}
    \item (Finite \LLR Variance) There exists $0<\Xi<\infty$ such that $\expValDist{\left(\log\frac{f_i^a(X)}{f_j^a(X)}\right)^2}{f_{i}^a} < \Xi$ for any $i, j\in\mathcal{H}$ and $a\in\mathcal{A}$.
    \label{assumption: finite LLR variance}
    \item (Positive Costs) For any action $a\in\mathcal{A}$, $C_a > 0$ with probability strictly greater than 0.
    \label{assumption: positive costs}
\end{enumerate}
The Separation assumption ensures that each action separates at least two hypotheses.
The Validity assumption ensures that all hypotheses are distinguishable.
The Finite \LLR Variance assumption, first introduced by Chernoff \cite{Chernoff1959SequentialHT}, implies that the \KLDstxt are finite, which in turn implies that under each action, the support is shared.
Note that the sample or cost supports may not be shared among different actions.

Although the Separation, Validity, and Positive Costs assumptions are not necessary, they ensure that the model is both simple and interesting.
Specifically, if the Separation assumption does not hold, some actions cannot distinguish between any hypotheses, and applying them is not only uninformative but also meaningless.
If the Validity assumption does not hold, then the indistinguishable hypotheses (under all available actions) can be treated as one until the \DM retires.
In this case, the \DM must randomly select one of them and declare it true, bounding the error probability away from zero.
If the Positive Costs assumption does not hold, i.e., there exists some zero cost action $a_0$, then the \DM can apply $a_0$ infinite times to reduce the hypotheses set to a set containing only hypotheses whose output distribution under $a_0$ has zero \TVDtxt from the output distribution under $\theta$ and $a_0$.
This reduced set can then be treated as $\mathcal{H}$.

Let $\Psi$ be the selection process generating the action sequence $\{A_n\}_{n=1}^\infty$, $N$ be the (possibly random) stopping time of $\Gamma$ and $\hat{\theta}$ be the decision at time $N$. 
An admissible strategy for the \CASHT, $\Gamma\triangleq(\Psi, \hat{\theta})$, is a strategy solving
\begin{align}
    \label{eq: original problem}
    \begin{split}
        \min_{\Gamma}\quad& \expValDist{\sum_{n=1}^{N} C_{A_n} \middle| \Gamma}{N, \indexedSet{C}{A_1}{A_N}}
        \\
        \mathrm{s. t.}\quad&    p_e(\Gamma)\leq \delta
    \end{split}
    ,
\end{align}
where $p_e(\Gamma)\triangleq\expVal{\mathbb{P}(\hat{\theta} \neq \theta) \middle| \Gamma}$ is the average error probability of policy $\Gamma$.

The problem in Eq. \eqref{eq: original problem} considers three key components of the problem: the error probability, $p_e$, the number of samples taken until termination, $N$, and the action costs, $\{C_a\}_{a\in\mathcal{A}}$.
The relationship between these components is not trivial; increasing the number of samples (and thus the expected total cost) can significantly reduce the error probability.
Conversely, by abstaining from sampling, the error probability increases (up to $1-1/H$), violating the constraint.
Accordingly, we must carefully consider how these components intertwine.

\section{Fixed Acquisition Costs Model}
\label{section: model and problem review}
We first consider minimizing the total expected cost in the simple case where all costs are deterministic (i.e., replacing $C_a$ with a fixed constant $c_a$) to obtain key insights into the design of the \CASHT algorithms.
Note that when the \DM never cancels its actions, it follows that $0 < c_a = \expVal{C_a}$ since the costs are statistically independent of $\theta$ or $\Gamma$.
However, if it does leverage its action canceling, $c_a$ may be evaluated to a different quantity (see Section \ref{section: impact action cancellation} below).

In the \CA settings, the expected total cost reflects not only how long the algorithm samples before stopping, but also which actions it chooses along the way.
In the following theorem, we prove that for any algorithm that bases its decisions on the collected evidence and terminates almost surely, the conditional (on $\theta$) expected total cost can be conveniently decomposed into the product of per-action average contribution and the expected number of samples, effectively separating sampling behavior from the cost model.
\begin{lemma}
    \label{lemma: expected total cost decomposition to expected product}
    Assume $0 < \expVal{N|\Gamma} < \infty$.
    Denote by $\myVec{\lambda}_{\theta, \Gamma}$ the action distribution under algorithm $\Gamma$ conditioned on $\theta$ being the true hypothesis.
    Then, the expected total cost can be decomposed into
    \begin{align*}
        \expVal{\sum_{n=1}^{N} c_{A_n} \middle | \Gamma}
        =
        \expValDist{
            \expValDist{c_{A}|\theta,\Gamma}{A\sim\myVec{\lambda}_{\theta, \Gamma}} \times \expVal{N|\theta, \Gamma}
        }{\theta}
        .
    \end{align*}
\end{lemma}
\begin{IEEEproof}
    \ifShowSupp
        See Appendix \ref{lemma: expected total cost decomposition to expected product proof}.
    \fi
\end{IEEEproof}
Lemma \ref{lemma: expected total cost decomposition to expected product} implies that when the action distribution remains fixed, one may effectively separate the effects of the cost from the number of samples \emph{as if they were independent}.

For the rest of the paper, to simplify notation, we drop the conditioning on $\Gamma$, and the discussed policies should be clear from context.
A useful way to interpret Lemma \ref{lemma: expected total cost decomposition to expected product} is that each algorithm induces an effective action distribution, determined by the expected fraction of samples collected from each action.
The expected cost per sample is then evaluated with respect to this induced distribution and can be interpreted as a single \emph{uniform acquisition cost} for all actions.
This cost is not directly affected by sample counts; rather, they determine the extent to which each action influences the overall expected cost.
For example, if an action is never selected, its expected sample count is zero, and the induced action distribution assigns it no mass, completely negating its effect on the expected cost.

Moreover, when the costs are independent of $\delta$, the expected total cost’s scaling laws in $\delta$ are completely determined by the expected number of samples’ scaling laws in $\delta$.
Namely, Lemma \ref{lemma: expected total cost decomposition to expected product} also established that \SHT algorithms retain their $\bigTheta{\log(1/\delta)}$ scaling laws in our model, but since the expected cost is not accounted for in their design, they can perform poorly in practice.

Note that when replacing $c_a = \delta$ for any action, as accepted in the literature (see, for example, \cite{Chernoff1959SequentialHT, Nitinawarat2013_SHT_Argmax2_KLD_wProofs, Cohen_Zhao2015_SHT_AnomalyDetection, Gafni2023_CompositeHT, Citron_Cohen_Zhao2024_DGF_on_Hidden_Markov_Chains}), the classic active \SHT formulation is obtained;
The expected total cost becomes $\delta \expVal{N}$, and the corresponding Lagrangian of the problem in Eq. \eqref{eq: original problem} is $\delta\expVal{N} + L\left( p_e - \delta \right)$ for some $L\geq0$, whose minimization is akin to minimizing the \ABR \cite{Naghshvar_Javidi2013_SHT_DynamicProgramming}.
Therefore, effective \CASHT algorithms must retain a sublinear expected number of samples in $1/\delta$ for vanishing \ABR as $\delta$ approaches zero.

We now shift our focus to the expected cost $\expValDist{c_{A}|\theta, \Gamma}{A}$.
Optimizing the expected cost is not merely a matter of favoring cheaper actions.
Cheap actions may be insufficiently informative to guarantee reliable detection, or may require so many samples to separate certain hypotheses that their total expected cost exceeds that of a more expensive but substantially more informative action.
Conversely, some actions may be highly favorable, combining low cost with strong discriminative power, while others may be unfavorable in both aspects.
A \CASHT algorithm should therefore balance informativeness and cost, favoring actions that provide reliable evidence at low acquisition cost and avoiding actions whose cost is not justified by their contribution to hypothesis separation.

To optimize the expected total cost, we combine the decomposition into a weighted sum of products in Lemma \ref{lemma: expected total cost decomposition to expected product} with the asymptotic bounds on $\expVal{N|\theta}$ in \cite[Lemma~8]{vershinin2025iterativehypothesispruningdistributionbased}, to obtain the following asymptotic characterization of the expected total cost under \MSPRT:
\begin{theorem}
    \label{theorem: expected total cost bounds}
    Denote $I_\theta(\myVec{\lambda}) \triangleq \min_{j\neq \theta} \expValDist{ \KLD{f_\theta^A}{f_j^A} }{A\sim \myVec{\lambda}}$, where $\myVec{\lambda}\in\probSimplexA$.
    Then, for any $\delta\in (0, 1)$,
    \begin{align*}
        \expVal{\sum_{n=1}^{N} c_{A_n} \middle | \theta, \myVec{\lambda}}
        =
        (1+\littleO{1})\frac{ \log(1/\delta) }{ I_\theta(\myVec{\lambda})/\expValDist{c_A|\theta}{A\sim\myVec{\lambda}} }
        ,
    \end{align*}
    where the $\littleO{\cdot}$ term is with respect to $\delta\to0$.
\end{theorem}
\begin{IEEEproof}
    In the proof of \cite[Lemma~8]{vershinin2025iterativehypothesispruningdistributionbased}, we showed that $g(\delta) = 2\delta+\xi_1/(\log(1/\delta))$ and $h(\delta)=\xi_2/(\log(1/\delta))$ (for some \LLR-related constants $\xi_1, \xi_2>0$ existing due to Assumption \ref{assumption: finite LLR variance}) satisfy $(1+h(\delta))\log (1/\delta)/I_{\theta}(\myVec{\lambda}) \geq \expVal{N|\theta}\geq (1-g(\delta))\log (1/\delta)/I_{\theta}(\myVec{\lambda})$.
    Notably, $h(\delta), g(\delta)\geq0$ and $h(\delta), g(\delta)\to0$ as $\delta\to0$.
    By multiplying all sides by $\expValDist{c_A|\theta}{A\sim\myVec{\lambda}} > 0$, we obtain the result.
\end{IEEEproof}
Theorem \ref{theorem: expected total cost bounds} establishes that the conditional expected total cost is asymptotically equivalent to the amount of information bits required to describe the desired error probability $\delta$, namely $\log(1/\delta)$, divided by the expected information gain per unit expected cost.
Nevertheless, both of the additive corrections can be sharpened beyond $\bigO{(\log(1/\delta))^{-1}}$.
For the lower bound, for example, a sharper additive correction of $\bigO{(\log(1/\delta))^{-1/3}}$ can be obtained as argued in \cite[Corollary~2]{Naghshvar_Javidi2013_SHT_DynamicProgramming}\footnote{More precisely, their Corollary 2 relies on the family of lower bounds in their Proposition 1 that essentially bounds the expected number of samples; see Eq. (80) in their supplementary material \cite{Naghshvar_Javidi2013_SHT_DynamicProgramming_Supp}.}.
These improvements remain out of scope for this paper.

A closer inspection of the proof of \cite[Proposition~1]{Naghshvar_Javidi2013_SHT_DynamicProgramming}, given in \cite{Naghshvar_Javidi2013_SHT_DynamicProgramming_Supp}, shows that the obtained asymptotic lower bound in the proof of Theorem \ref{theorem: expected total cost bounds} applies to \emph{any algorithm} whose stopping rule guarantees $p_e\leq \delta$.
This, combined with the upper bound in the proof of Theorem \ref{theorem: expected total cost bounds}, implies that \MSPRT-based algorithms are essentially optimal for any given action distribution.
However, Theorem \ref{theorem: expected total cost bounds} relies on the knowledge of the true hypothesis $\theta$.
Accordingly, we will refer to the result in Theorem \ref{theorem: expected total cost bounds} as an \emph{oracle} that knows the true hypothesis, but still requires drawing samples (using the given action distribution) until the desired confidence level is achieved.

Naturally, the expected total cost obtained by the oracle can be optimized.
Particularly, the leading constant can be optimized as a function of $\myVec{\lambda}$, leading to the following design principle for \CASHT algorithms: actions should be selected to maximize the ratio of expected information gain to expected cost, as induced by the \emph{conditional action distribution} under each true hypothesis $\theta$.
That is, 
\begin{corollary}[\CASHT Design Principle]
    \label{corollary: expected information gain per expected cost optimization goal}
    For any $\theta\in\mathcal{H}$, the asymptotic optimal action distribution to minimize the expected total cost is the solution to the following \LP problem:
    \begin{align*}
        \max_{\myVec{\lambda}_\theta\in\probSimplexA}
            \frac{ \min_{j\neq \theta}\expValDist{ \KLD{f_\theta^A}{f_j^A} }{A\sim \myVec{\lambda}_\theta} }
                 { \expValDist{c_A}{A\sim\myVec{\lambda}_\theta} }
        .
    \end{align*}
\end{corollary}
\textit{Remark:} Since $c_a > 0$ for any $a\in\mathcal{A}$, the above is always finite.
Since $\probSimplexA$ is compact, the maximum exists from continuity in $\myVec{\lambda}$.
Thus, Assumption \ref{assumption: validity} ensures that the above is positive, which in turn implies that $I_\theta > 0$ at the maximizing action distribution.
To see this, observe that $\myVec{1}/\abs{\mathcal{A}}$ assigns positive mass to actions separating $j\neq \theta$ and is feasible.

We say that a \CASHT algorithm is \emph{asymptotically optimal} in $\delta$ if it satisfies all of the following conditions.
\begin{enumerate}[label=(B\arabic*)]
    \item (Satisfiability) The prescribed error constraint is satisfied, i.e., $p_e\leq \delta$.
    \label{asymptotic optimality definition: satisfiability}
    \item (Optimal Scaling in $\delta$) Its expected total cost scales like $\bigO{\log(1/\delta)}$.
    \label{asymptotic optimality definition: optimal scaling}
    \item (Correct Scaling Constant) The ratio between its expected total cost and that of an oracle selecting actions according to Corollary \ref{corollary: expected information gain per expected cost optimization goal} converges to 1 as $\delta\to0$.
    \label{asymptotic optimality definition: scaling constant}
\end{enumerate}
The above conditions capture increasing notions of optimality.
Satisfiability alone guarantees that the algorithm is \emph{feasible}.
Optimal Scaling in $\delta$ establishes \emph{order-optimality} and guarantees vanishing \ABR in the classic settings, while Correct Scaling Constant ensures that the leading-order constant is the smallest possible in our non-homogeneous cost model.
We now shift our focus to improving order-optimal \SHT algorithms to achieve asymptotic optimality for \CASHT.

\section{Adapting \texorpdfstring{ \SHT Algorithms to \CASHT }{ SHT Algorithms to CASHT } }
\label{section: Policies}
Using the key observation that all \CASHT algorithms should optimize the expected information gained per expected cost, we adapt the Chernoff scheme \cite{Chernoff1959SequentialHT} and NJ1 \cite{Naghshvar_Javidi2013_SHT_DynamicProgramming} to \CASHT in this section.

\subsection{The Adapted Chernoff Scheme}
\label{subsection: adapting Chernoff}
The classic Chernoff scheme is an asymptotically optimal (for \SHT) stochastic algorithm in which actions at each time step are chosen randomly, and the hypothesis with the highest posterior probability guides the action distribution.
Once the posterior probability of at least one hypothesis exceeds $1-\delta$, the procedure terminates, and the \DM declares the hypothesis with the largest posterior as the true hypothesis.

The action guiding distribution for $\hII$, i.e., the distribution from which actions are drawn when $\hII$ is believed to be true, is the one optimizing $I_i(\myVec{\lambda})$.
Namely, it is the solution of the following \LP problem:
\begin{align}
    \label{eq: Chernoff Policy Dist}
    \myVec{\lambda}_i^{\mathrm{CH}}
    =
    \argmax{\min_{j\neq i}\expValDist{ \KLD{f_i^A}{f_j^A} }{A\sim \myVec{\lambda}_i}}{ \myVec{\lambda}_i\in\probSimplexA }
    .
\end{align}
\textit{Remarks}: (i) $\{\myVec{\lambda}_h^{\mathrm{CH}}\}_{h\in\mathcal{H}}$ are only computed once and before the \DM begins operating.
(ii) The stopping rule of some posterior exceeding $1-\delta$ for the first time is equivalent to stopping rule in \MSPRT where some hypothesis has its \LLR exceed $\log(1/\delta)$ for the first time \cite{Dragalin_etAl_1999_MSPRT_AsympOpt}.

The \CA variant of the Chernoff scheme, which we term \CACH, is similar to the classic Chernoff, differing only in its guiding distribution, adjusted according to Corollary \ref{corollary: expected information gain per expected cost optimization goal}: 
\begin{align}
    \label{eq: Chernoff Policy Dist With Costs}
    \myVec{\lambda}_i^{\mathrm{CA-CH}}
    =
    \argmax{
        \frac{ \min_{j\neq i}\expValDist{ \KLD{f_i^A}{f_j^A} }{A\sim \myVec{\lambda}_i} }
             { \expValDist{c_A}{A\sim\myVec{\lambda}_i} }
    }{ \myVec{\lambda}_i\in\probSimplexA }
    .
\end{align}
As in the original Chernoff scheme, $\{\myVec{\lambda}_h^{\mathrm{CA-CH}}\}_{h\in\mathcal{H}}$ are computed only once in an offline stage.
These distributions can be computed using \LP as well:
\begin{proposition}
    \label{proposition: adjusted Chernoff dist}
    For any $i\in\mathcal{H}$, let $\myVec{d}_{ij} = \{\KLD{f_i^a}{f_j^a}\}_{a\in\mathcal{A}}$ be a column vector containing the \KLDtxt between $f_i^a$ and $f_j^a$ for any $a\in\mathcal{A}$.
    Let $\myVec{c} = \{c_a\}_{a\in\mathcal{A}}$ be a column vector containing all costs associated with the actions.
    Then, $\myVec{\lambda}_i^{\mathrm{CA-CH}} = \myVec{y}^*/z^*$ where the $\myVec{y}^*$ and $z^*$ are the solutions of:
    \begin{equation}
        \nonumber
        \begin{split}
            \max_{ \myVec{y}\in\mathbb{R}^{\abs{\mathcal{A}}}, z\in\mathbb{R} }\quad&
                \min_{j\neq i} \myVec{d}_{ij}^T\myVec{y}
            \\
            \mathrm{s. t.}\quad&    
                \begin{cases}
                    \myVec{1}^T\myVec{y} -z = 0, &\myVec{c}^T\myVec{y} = 1 \\
                    \myVec{y} \geq \myVec{0}, &z\geq 0
                \end{cases}
        \end{split}
    \end{equation}
\end{proposition}
\begin{IEEEproof}
    The problem in Eq. \eqref{eq: Chernoff Policy Dist With Costs} can be transformed into the above optimization problem in a similar way that Linear-Fractional Programming problems can be transformed to \LP problems, cf. \cite[Chapter~4.3.2]{boyd_vandenberghe_2004}.
    \ifShowSupp
        See Appendix \ref{proposition: adjusted Chernoff dist proof} for details.
    \fi
    Notably, our remark on Corollary \ref{corollary: expected information gain per expected cost optimization goal} is also reflected here;
    $\{\myVec{\lambda}_h^{\mathrm{CA-CH}}\}_{h\in\mathcal{H}}$ are always well-defined since $z = 0$ is always infeasible.
    Particularly, if $z=0$ then $\myVec{1}^T\myVec{y} = 0$, which in turn implies that $\myVec{y}=\myVec{0}$, violating $\myVec{c}^T\myVec{y} = 1$.
\end{IEEEproof}
We now formally describe how \CACH operates.
Denote by $\myVec{\rho}_n\in\probSimplex{H}$ the posterior probability over the hypotheses after taking the $n^{\mathrm{th}}$ sample $x_n$.
That is, $\vectorComponent{\rho}{i}[n] = \eta_n\times\vectorComponent{\rho}{i}[n-1]\times f_i^a(x_n)$ where $\eta_n \triangleq 1/(\sum_{h\in\mathcal{H}} \vectorComponent{\rho}{h}[n]\times f_h^a(x_n))$ is the normalization factor and $\myVec{\rho}_0\triangleq \myVec{1}/H$ is the uniform prior.
For any $n\geq1$, the \DM draws action $A_n\sim \myVec{\lambda}_{i_n^*}^{\mathrm{CA-CH}}$ where $i_n^* \triangleq \argmax{ \vectorComponent{\rho}{h}[n-1] }{ h\in\mathcal{H} }$ is the most likely hypothesis at timestep $n$ (with ties broken arbitrarily), obtains a sample, and updates posteriors until the first time some $i\in\mathcal{H}$ has $\vectorComponent{\rho}{i}[n]\geq 1-\delta$, which is declared as true.

The following summarizes the properties of \CACH: 
\begin{theorem}
    \label{theorem: Chernoff order optimality}
    \CACH is asymptotically optimal, i.e., it satisfies conditions \ref{asymptotic optimality definition: satisfiability}, \ref{asymptotic optimality definition: optimal scaling}, and \ref{asymptotic optimality definition: scaling constant}.
\end{theorem}
\begin{IEEEproof}
    Since \CACH follows the same procedure as the classic Chernoff scheme, its error probability does not exceed $\delta$ (e.g., \cite[Lemma~3]{Chernoff1959SequentialHT}), establishing condition \ref{asymptotic optimality definition: satisfiability}.
    To show that conditions \ref{asymptotic optimality definition: optimal scaling} and \ref{asymptotic optimality definition: scaling constant} hold, decompose the expected total cost:
    \begin{align}
        \label{eq: good-bad decomposition}
        \sum_{n=1}^{N} c_{A_n}
        =
        \sum_{n=1}^{N} c_{A_n}\indicator{ i_n^* = \theta }
        +
        \sum_{n=1}^{N} c_{A_n}\indicator{ i_n^* \neq \theta }
        ,
    \end{align}
    where $\indicator{A}$ is the indicator function of event $A$.
    The former sum corresponds to the timesteps in which \CACH follows the footsteps of the oracle, and we are left to show this sum, in expectation, scales like $\bigO{\log(1/\delta)}$ (condition \ref{asymptotic optimality definition: optimal scaling}), and is much larger than the latter (condition \ref{asymptotic optimality definition: scaling constant}).

    To show these, we will leverage the accumulated \LLR between two hypotheses $i$ and $j$, defined as
    \begin{align*}
        L_{ij}(\indexedSet{x}{1}{n}; \indexedSet{a}{1}{n})
        \triangleq
        \sum_{t=1}^n \log \frac{f_i^{a_t}(x_t)}{f_j^{a_t}(x_t)}
        .
    \end{align*}
    Denote by $\mathcal{F}_n$ the $\sigma$-algebra generated by $\indexedSet{X}{1}{n}$ and $\indexedSet{A}{1}{n}$ (and policy $\Gamma$).
    Denote by $D_{i,k}$ the drift of the accumulated \LLR between $\theta$ and $k$ when taking actions under the belief that $i$ is the true hypothesis.
    Namely, $D_{i,k}\triangleq \expValDist{ \KLD{f_{\theta}^A}{f_k^A} }{A\sim \myVec{\lambda}_{i}^{\mathrm{CA-CH}}}$.
    Note that $D_{\theta, k} = I_\theta(\myVec{\lambda}_{\theta}^{\mathrm{CA-CH}})$.
    Decompose:
    \begin{align*}
        L_{\theta, k}
        &=
        \sum_{t=1}^n \left(\log \frac{f_\theta^{A_t}}{f_k^{A_t}} - D_{i_t^*, k}\right)
        + \sum_{t=1}^n D_{i_t^*, k}
        .
    \end{align*}
    Observe that $M_t (k) \triangleq \log ((f_\theta^{A_t}(X_t)/(f_k^{A_t}(X_t))) - D_{i_t^*, k}$ have zero expectation.
    Denote by $J_i(n) \triangleq \sum_{t=1}^n \indicator{ i_t^* = i }$ the number of times hypothesis $i$ was believed to be true until timestep $n$ (inclusive).
    Let $\mathcal{W}(n) \triangleq \sum_{j\neq\theta}J_j(n)$ be the number of times $\theta$ was not believed to be true until timestep $n$.
    Note that $J_\theta(n) = n - \mathcal{W}(n)$.
    
    We bound the drift sum from below $\sum_{t=1}^n D_{i_t^*, k}\geq I_\theta(\myVec{\lambda}_{\theta}^{\mathrm{CA-CH}}) \times (n-\mathcal{W}(n))$ for any $k\neq\theta$.
    
    At the stopping time $N$, the accumulated \LLR satisfies:
    \begin{align*}
        \log \frac{1-\delta}{\delta} \leq \abs{\max_{k\neq\theta} L_{\theta, k}} \leq \log \frac{1-\delta}{\delta} + \max_{k\neq\theta}\sigma_{\theta k}(a_N)
    \end{align*}
    where $\sigma_{\theta k}(a_N)\geq 0$ is an \LLR (between $\theta$ and $k$) overshoot term depending on the last acquired sample via the last action, $a_N$, and $\log((1-\delta)/\delta)$ is the \LLR-equivalent of the $1-\delta$ posterior stopping \cite[Section~II.B]{Dragalin_etAl_1999_MSPRT_AsympOpt}.
    In particular, since $\log \frac{1-\delta}{\delta} = \log(1/\delta) + \log(1-\delta) = \log(1/\delta) + \littleO{1}$, we have
    \begin{align}
        \nonumber
        &\log\frac{1}{\delta} + \littleO{1} + \max_{k\neq\theta}\sigma_{\theta k}(a_N)
        \geq
        \min_{k\neq\theta} L_{\theta, k}
        \\
        \nonumber
        &\quad\geq
        \min_{k\neq\theta} \sum_{t=1}^N M_t + I_\theta(\myVec{\lambda}_{\theta}^{\mathrm{CA-CH}}) \times (N-\mathcal{W}(N))
        \\\nonumber
        &\quad\geq
        N\times\bigg(
            \frac{\min_{k\neq\theta} \sum_{t=1}^N M_t}{N} + I_\theta(\myVec{\lambda}_{\theta}^{\mathrm{CA-CH}})
            \\\label{eq: LLR lower bound}
            &\quad\quad\quad\quad\quad\quad\quad\quad
            -I_\theta(\myVec{\lambda}_{\theta}^{\mathrm{CA-CH}})\times\frac{\mathcal{W}(N)}{N}
        \bigg)
        .
    \end{align}
    We now need to show that Eq. \eqref{eq: LLR lower bound} can be integrated, i.e., that $N<\infty$ almost surely.
    To this end, we claim the following:
    \begin{proposition}
        \label{proposition: sum Mn = o(n)}
        $\lim_{n\to\infty}(\sum_{t=1}^n M_t)/n = 0$ almost surely.
    \end{proposition}
    \begin{IEEEproof}
        \ifShowSupp
            See Appendix \ref{proposition: sum Mn = o(n) proof}.
        \fi
    \end{IEEEproof}
    \begin{proposition}
        \label{proposition: Jj = o(n)}
        For any $k\neq\theta$, $J_k(n) = \littleO{n}$.
        Particularly, $\lim_{n\to\infty} J_k(n)/n = 0$.
    \end{proposition}
    \begin{IEEEproof}
        \ifShowSupp
            See Appendix \ref{proposition: Jj = o(n) proof}.
        \fi
    \end{IEEEproof}
    Note that $\mathcal{W}(n) = (H-1)\times\littleO{n} = \littleO{n}$ with respect to $\delta$, but grows linearly in $H$.
    
    Combining both limits in these propositions yields that $\min_{k\neq\theta} L_{\theta, k}\geq I_\theta\times N \times(1+\littleO{1})$, which grows as $N\to\infty$, establishing that the accumulated \LLR grows in favor of $\theta$ as evidence is accumulated, i.e., $N<\infty$ almost surely.

    Since $\{\sigma_{\theta k}(a_N)\}_k$ are $\delta$-independent finite-expectation constants (see Lorden’s Inequality \cite{lorden1970excess}), Eq. \eqref{eq: LLR lower bound} implies
    \begin{align}
        \label{eq: samples bound CACH}
        \expVal{N|\theta}
        \leq
        (1+\littleO{1})\frac{\log(1/\delta)}{I_\theta(\myVec{\lambda}_{\theta}^{\mathrm{CA-CH}})} + \expVal{\mathcal{W}(N)}
        .
    \end{align}

    Multiplying Eq. \eqref{eq: samples bound CACH} by $\expValDist{c_A}{A\sim \myVec{\lambda}_{\theta}^{\mathrm{CA-CH}} }$ yields a bound on a single addend as in Lemma \ref{lemma: expected total cost decomposition to expected product}.
    Averaging over $\theta$ gives us a direct upper bound on Eq. \eqref{eq: good-bad decomposition}.
    Specifically, the left addend in Eq. \eqref{eq: samples bound CACH} bounds the left sum in Eq. \eqref{eq: good-bad decomposition} and reflects the oracle result in Theorem \ref{theorem: expected total cost bounds}.
    Therefore, we are left to show that $\expVal{\mathcal{W}(N)}$ does not grow in $\delta$.
    This is given in the following lemma that, essentially, establishes both \ref{asymptotic optimality definition: optimal scaling} and \ref{asymptotic optimality definition: scaling constant}:
    \begin{lemma}
        \label{lemma: E[W(N)] is bounded}
        There exists some constant $\iota > 0$ such that $\expVal{\mathcal{W}(N)} < \iota$ for any $\delta$. 
    \end{lemma}
    \begin{IEEEproof}
        The proof follows by using the sum-tail identity for expectation and the union bound (over hypotheses $k\neq\theta$) to show that the convergences of the probability series (over timesteps $n$) of $\prob{ \exists m\geq n: L_{\theta, k}\leq 0 }$ defines an upper bound of $\expVal{\mathcal{W}(N)}$, i.e., $\iota$.
        Then, we proceed to show that these probabilities are exponentially bounded, namely, that there are some $\xi, \zeta>0$ such that the above probability is bounded by $\xi \times 2^{-\zeta n}$.
        This is done by studying the concentration of the tilted accumulated \LLR, which is shown to be a random walk with positive drift.
        \ifShowSupp
            See Appendix \ref{lemma: E[W(N)] is bounded proof} for details.
        \fi
    \end{IEEEproof}
\end{IEEEproof}
The key limitation of the \CACH scheme is the potential instability of its action-selection rule when the accumulated samples do not yet provide sufficient discrimination among the hypotheses, leading to oscillations in $i_n^*$, which may degrade performance in some scenarios.
To mitigate this behavior, the algorithm in the sequel introduces an explicit exploration phase that may slightly degrade performance in some scenarios but guarantees persistent discrimination regardless of the current hypothesis estimate $i_n^*$, thereby ensuring sufficiently rapid convergence to the true hypothesis.

\subsection{The Adapted NJ1 Algorithm}
\label{subsection: adapting NJ1}
Similar to the Chernoff scheme, the classic NJ1 algorithm is an asymptotically optimal stochastic algorithm for \SHT in which actions are drawn randomly from a distribution guided by the current belief about the true hypothesis.
In fact, the guiding distributions in NJ1 are precisely those that guide action selection in the Chernoff scheme.
However, unlike the Chernoff scheme, NJ1 includes an exploration phase that persists until some hypothesis has a posterior probability exceeding the user-tuned \emph{fixed} threshold $\tilde{\rho} > 0.5$.
The action-guiding distribution in the exploration phase emphasizes the worst-case separation and is obtained by solving
\begin{align}
    \label{eq: NJ1 Policy Exploration Dist}
    \myVec{\lambda}_{\mathrm{Exp}}^{\mathrm{NJ1}}
    =
    \argmax{ \min_{i\in\mathcal{H}} \min_{j\neq i}\expValDist{ \KLD{f_i^A}{f_j^A} }{A\sim \myVec{\lambda}}}{ \myVec{\lambda}\in\probSimplexA }
    .
\end{align}
Similar to the Chernoff scheme, the adapted NJ1, termed \CANJ, requires adjusting the guiding distributions in both phases.
The guiding distributions during the exploitation phase, i.e., when it operates as the \CACH scheme, are $\{\myVec{\lambda}_h^{\mathrm{CA-CH}}\}_{h\in\mathcal{H}}$.
For the exploration phase, the worst-case separation per mean cost, i.e., $\min_{h\in\mathcal{H}} \{I_h(\myVec{\lambda})/\expValDist{c_{A}}{A\sim\myVec{\lambda}}\}$ is optimized:
\begin{align}
    \label{eq: NJ1 Exploration Dist With Costs}
    \myVec{\lambda}_{\mathrm{Exp}}^{\mathrm{CA-NJ1}}
    =
    \argmax{
        \min_{h\in\mathcal{H}}
        \frac{ \min_{j\neq h}\expValDist{ \KLD{f_h^A}{f_j^A} }{A\sim \myVec{\lambda}} }
             { \expValDist{c_{A}}{A\sim\myVec{\lambda}} }
    }{ \myVec{\lambda}\in\probSimplexA }
    .
\end{align}
As before, $\myVec{\lambda}_{\mathrm{Exp}}^{\mathrm{CA-NJ1}}$ is computed once in an offline stage, and can be computed via \LP similar to $\{\myVec{\lambda}_h^{\mathrm{CA-CH}}\}_{h\in\mathcal{H}}$:
\begin{proposition}    
    \label{proposition: adjusted NJ1 exploration dist}
    For any $i\in\mathcal{H}$, let $\myVec{d}_{ij} = \{\KLD{f_i^a}{f_j^a}\}_{a\in\mathcal{A}}$ be a column vector containing the \KLDtxt between $f_i^a$ and $f_j^a$ for any $a\in\mathcal{A}$.
    Let $\myVec{c} = \{c_a\}_{a\in\mathcal{A}}$ be a column vector containing all costs associated with the actions.
    Then, $\myVec{\lambda}_{\mathrm{Exp}}^{\mathrm{CA-NJ1}} = \myVec{y}^*/z^*$ where the pair $\myVec{y}^*$ and $z^*$ are the solutions of:
    \begin{equation}
        \nonumber
        \begin{split}
            \max_{ \myVec{y}\in\mathbb{R}^{\abs{\mathcal{A}}}, z\in\mathbb{R} }\quad&
                \min_{i\in\mathcal{H}}\min_{j\neq i} \myVec{d}_{ij}^T\myVec{y}
            \\
            \mathrm{s. t.}\quad&    
                \begin{cases}
                    \myVec{1}^T\myVec{y} -z = 0, &\myVec{c}^T\myVec{y} = 1 \\
                    \myVec{y} \geq \myVec{0}, &z\geq 0
                \end{cases}
        \end{split}
    \end{equation}
\end{proposition}
\begin{IEEEproof}
    The proof is identical to the proof of Proposition \ref{proposition: adjusted Chernoff dist}.
    \ifShowSupp
        See Appendix \ref{proposition: adjusted NJ1 exploration dist proof} for details.
    \fi    
\end{IEEEproof}
Formally, \CANJ operates as follows.
Fixes $\tilde{\rho} > 1/2$.
Then, at each timestep $n\geq1$, the \DM draws action $A_n\sim \myVec{\lambda}_{i_n^*}^{\mathrm{CA-CH}}$ if $\vectorComponent{\rho}{i_n^*}[n-1] \geq \tilde{\rho}$ or $A_n\sim \myVec{\lambda}_{\mathrm{Exp}}^{\mathrm{CA-NJ1}}$ if $\vectorComponent{\rho}{i_n^*}[n-1] < \tilde{\rho}$, obtains a sample, and updates posteriors until the first time some $i\in\mathcal{H}$ has $\vectorComponent{\rho}{i}[n]\geq 1-\delta$, which is declared as true.

We now argue that \CANJ has similar properties to \CACH using similar arguments as in the proof of Theorem \ref{theorem: Chernoff order optimality} in the following:
\begin{theorem}
    \label{theorem: NJ1 order optimality}
    \CANJ is asymptotically optimal.
\end{theorem}
\begin{IEEEproof}
    Condition \ref{asymptotic optimality definition: satisfiability} follows similarly to the classic NJ1.
    Similar to the proof of Theorem \ref{theorem: Chernoff order optimality}, we decompose the expected total cost into three components corresponding to correct exploitations (corresponding to the samples taken while following the oracle), erroneous exploitations, and exploration:
    \begin{align}
        \nonumber
        \sum_{n=1}^{N} c_{A_n}
        &=
        \sum_{n=1}^{N} c_{A_n}\indicator{ i_n^* = \theta, \text{exploit at timestep $n$} }
        \\\nonumber
        &\quad
        +
        \sum_{n=1}^{N} c_{A_n}\indicator{ i_n^* \neq \theta, \text{exploit at timestep $n$} }
        \\
        \label{eq: good-bad-explore decomposition}
        &\quad
        +
        \sum_{n=1}^{N} c_{A_n}\indicator{ \text{explore at timestep $n$} }        
        .
    \end{align}
    To compute the expectation, i.e., integrate Eq. \eqref{eq: good-bad-explore decomposition}, we need to show that the stopping time of \CANJ, $N$, the timestep from which \CANJ always draws actions according to $\myVec{\lambda}_{\theta}^{\mathrm{CA-CH}}$, $N_0=\sup\{n\in\mathbb{N}: i_n^*\neq \theta\}$, and the timestep in which the exploration phase expires are all finite almost surely.
    To show this, decompose the \LLR as:
    \begin{align*}
        &\left(\sum_{t=1}^n M_t + \sum_{t=1}^n D_{i_t^*, k}\right) \indicator{ \text{exploit at timestep $t$} }
        \\
        &\quad
        +
        \left(\sum_{t=1}^n M_t + \sum_{t=1}^n D_{\mathrm{Explore}}\right)\indicator{ \text{explore at timestep $t$} }
        ,
    \end{align*}
    where $D_{i_t^*, k}$ and $D_{\mathrm{Explore}} = \min_{h\in\mathcal{H}} I_h (\myVec{\lambda}_{\mathrm{Exp}}^{\mathrm{CA-NJ1}}) > 0$ are the \LLR drifts at timestep $t$ or during exploration, respectively.
    The sum corresponding to the exploitation phase is the same in the proof of Theorem \ref{theorem: Chernoff order optimality}, which establishes that $N_0, N <\infty$ almost surely.
    Accordingly, our focus shifts to the sum corresponding to the exploration phase.

    Conditioned on the samples, we have \cite[Section~II.B]{Dragalin_etAl_1999_MSPRT_AsympOpt}
    \begin{align*}
        \vectorComponent{\rho}{\theta}[n]
        =
        ( 1 + \sum_{j\neq \theta}2^{-L_{\theta,j}} )^{-1}
        .
    \end{align*}
    Let $M_0 = \sup\{ n\in\mathbb{N} : \vectorComponent{\rho}{i_n^*}[n] < \tilde{\rho} \}$.
    Notably, the definition of $M_0$ assumes that both the exploration and algorithm stopping rules are disregarded.
    Additionally, $M_0$ bounds the end of the exploration phase in practice.
    
    Since $\vectorComponent{\rho}{i_n^*}[n] < \tilde{\rho}$ in exploration, we have $(1-\tilde{\rho})/\tilde{\rho} < \sum_{j\neq \theta}2^{-L_{\theta, j}}$, which implies that $(1-\tilde{\rho})/((H-1)\tilde{\rho}) < \max_{j\neq \theta}2^{-L_{\theta, j}}$.
    By taking logarithms, we have $\min_{j\neq \theta} L_{\theta, j} < \log ( (H-1)\tilde{\rho}/(1-\tilde{\rho}) ) \triangleq \zeta$, with $\zeta > 0$ as $\tilde{\rho} > 0.5$.
    Consequently, $\{\min_{j\neq \theta} L_{\theta, j} \leq 0\} \subset \{\min_{j\neq \theta} L_{\theta, j} < \zeta\}$.
    
    In other words, similar to $N_0$ and $N$, having $M_0 < \infty$ almost surely is synonymous with having $\min_{j\neq \theta} L_{\theta, j}$ tending to infinity almost surely.
    Unlike $N_0$ or $N$, however, to have $M_0 < \infty$ almost surely, $\min_{j\neq \theta} L_{\theta, j}$ must be able to tend to infinity from the exploration phase alone.
    Since $D_{\mathrm{Explore}}>0$, there exists some $\xi > 0$ such that $L_{\theta, k} \geq \sum_{t=1}^n M_t +\xi J_\theta(n) = \sum_{t=1}^n M_t +\xi \times n (1+\littleO{1}) = n\times ((\sum_{t=1}^n M_t)/n +\xi + \littleO{1})$, which tends to infinity, and $M_0 < \infty$ almost surely.

    We have established that (i) the exploration phase expires almost surely, (ii) \CANJ eventually takes actions as the oracle, and (iii) \CANJ almost surely stops.
    Now we can bound the number of samples of \CANJ as in the proof of Theorem \ref{theorem: Chernoff order optimality}.
    Here, if $M_\delta$ is the last timestep in which \CANJ explored up to termination, we have $M_\delta \leq M_0$ and the rest of the $N-M_\delta$ timesteps follow the same procedure as \CACH.
    Thus, in a similar computation separating the exploration and exploitation phases, we obtain that:
    \begin{align}
        \label{eq: samples bound CANJ exploit}
        \expVal{N-M_\delta|\theta} &
        \leq
        (1+\littleO{1})\frac{\log(1/\delta)}{I_\theta(\myVec{\lambda}_{\theta}^{\mathrm{CA-CH}})} + \expVal{\mathcal{W}(N)}
        \\
        \label{eq: samples bound CANJ explore}
        \expVal{M_\delta|\theta} &\leq
        \frac{\log ((1-\tilde{\rho})/\tilde{\rho}) }{\min_{h\in\mathcal{H}} I_h (\myVec{\lambda}_{\mathrm{Exp}}^{\mathrm{CA-NJ1}})}
        +
        \sigma_{\Xi}
        ,        
    \end{align}
    where $\sigma_{\Xi}$ is a constant corresponding to the overshoot term when transitioning from exploration to exploitation.
    By multiplying Eq. \eqref{eq: samples bound CANJ explore} by $\expValDist{c_A}{A\sim \myVec{\lambda}_{\mathrm{Exp}}^{\mathrm{CA-NJ1}} }$ and Eq. \eqref{eq: samples bound CANJ exploit} by $\expValDist{c_A}{A\sim \myVec{\lambda}_{\theta}^{\mathrm{CA-CH}} }$, and summing both, we obtain bounds corresponding to Eq. \eqref{eq: good-bad-explore decomposition}.
    Note that the overshoot term can be put into the $\littleO{1}$ of Eq. \eqref{eq: samples bound CANJ exploit} after summing.

    When $\tilde{\rho} > 0.5$ is independent of $\delta$, the right-hand expression in Eq. \eqref{eq: samples bound CANJ explore} is a constant.
    Since Eq. \eqref{eq: samples bound CANJ exploit} is identical to Eq.\eqref{eq: samples bound CACH}, conditions \ref{asymptotic optimality definition: optimal scaling} and \ref{asymptotic optimality definition: scaling constant} are established.
\end{IEEEproof}
We re-emphasize that the main gain from the exploration phase lies in tighter bounds on the error probabilities.
Specifically, in \CANJ, erroneous explorations are associated with the event $\{L_{\theta,k}\leq -\log (\tilde{\rho}/(1-\tilde{\rho}))\}$, whose probability is significantly lower than erroneous explorations associated with $\{L_{\theta,k}\leq 0\}$ in \CACH.

\section{Impact of Action Cancellation}
\label{section: impact action cancellation}
In this section, we extend the constant-cost theory to the random-cost model and then examine how action cancellation modifies the effective cost.

When $\expVal{C_a} < \infty$ for any $a$ and $\{C_a\}_{a\in\mathcal{A}}$ are independent of $\theta$ and $\Gamma$, the objective in Eq. \eqref{eq: original problem} can be simplified to:
\begin{align*}
    \expVal{\sum_{n=1}^{N} \expVal{C_{A_n}} }
    .
\end{align*}
Accordingly, the conditional expected total cost can be decomposed as in Lemma \ref{lemma: expected total cost decomposition to expected product} by replacing $c_a$ with $\expVal{C_a}$:
\begin{align}
    \label{eq: expected total cost decomposition 2}
    \expVal{\sum_{n=1}^{N} \expVal{C_{A_n}} \middle | \theta }
    =
    \expVal{N | \theta}
    \times
     \expValDist{ \expValDist{C_{A}}{C_{A}} | \theta}{A}
    .
\end{align}
Thus, the asymptotic optimality of the \CA algorithms with respect to the random cost model is preserved \emph{regardless} of when the realized cost is revealed.

Still, the realized costs and their expectations can be significantly large.
For example, tracking the health of network nodes by measuring response times to health-checking messages is orders of magnitude faster than waiting for control messages to be flooded across the network due to a timeout.
In both cases, the network controller idles for a long time waiting for a sample if a node goes down.
Namely, either the response time grows large, or a timeout occurs, typically taking tens or hundreds of seconds.
Both latencies should be compared with millisecond latencies in local-area networks.

\subsection{Impact on Ex-Post Cost Model}
\label{subsection: per-action deadline impact ex-post}
In the ex-post cost model, the realized cost of an action $A_n$ is revealed to the \DM only after the sample has been obtained, e.g., as in billing.
Since the \DM cannot cancel an action before observing the sample $X_n$, action cancellation has no operational effect as every selected action always consumes its full realized cost.
Namely, action cancellation cannot alter the expected total cost.
Accordingly, the \CA designs in the ex-post cost model should follow the design principles proposed in Section \ref{section: model and problem review}, where the fixed action costs are their means.

\subsection{Impact on Ex-Ante Cost Model}
\label{subsection: per-action deadline impact ex-ante}
In the ex-ante cost model, the realized cost of an action $A_n$ is revealed to the \DM before the sample is obtained, and, based on its realization, the \DM can cancel its in-progress action to obtain no sample (indicated by obtaining the $\abstain$ symbol).

To leverage the preemption ability in this model, we associate each action $a$ with a \textit{per-action deadline}, $T_a$.
Selecting $\{T_a\}_{a\in\mathcal{A}}$ will be discussed later, but sensible choices of $T_a$ must obey $0 < F_{C_a}(T_a) \leq 1$ for any $a$ (otherwise, the action will not yield samples).
Formally, applying action $A_n$ yields
\begin{align*}
    Y_n
    =
    \begin{cases}
        X_n & C_{A_n}\leq T_{A_n}
        \\
        \abstain & C_{A_n} > T_{A_n}
    \end{cases}
    ,
\end{align*}
whose associated cost is $\min\{C_a, T_a\}$.
In the latency interpretation, if the cost exceeds $T_a$, we abort at time $T_a$ and incur a cost of $T_a$, but no sample is obtained.
Accordingly, we decompose the number of times each action $a$ is applied, denoted by $N_a$, into two addends; the first is $N_{eff}^a$, which counts the number of samples acquired, and the second is $N_{cancel}^a$, which counts the number of times action $a$ has been canceled.
That is, $N_a = N_{eff}^a + N_{cancel}^a$.

Since $N_{eff}^a$ is the number of samples used to update the accumulated \LLRs or posteriors, it is the studied number of samples in the literature.
Consequently, by leveraging Wald’s Identity \cite[Proposition~2.18]{siegmund2013sequential}, we can compute $N_a$ as a function of $N_{eff}^a$:
\begin{proposition}
    \label{proposition: Na = f(Na_eff)}
    $\expVal{N_a|\theta} = \expVal{N_{eff}^a|\theta} / F_{C_a}(T_a)$
\end{proposition}
\begin{IEEEproof}
    \ifShowSupp
        See Appendix \ref{proposition: Na = f(Na_eff) proof}.
    \fi   
\end{IEEEproof}
\textit{Remark}: $1/F_{C_a}(T_a)$ is the factor increasing the number of times action $a$ is applied compared to the scenario when no action cancellation is used.
Specifically, when no action cancellation is used, the \DM sets $T_a=\infty$ and $1/F_{C_a}(T_a) = 1$.

Now, we move to study the objective function.
Conditioned on $\theta$, for any action sequence $\{a_n\}_{n=1}^{N}$, the objective becomes:
\begin{align}
    \nonumber
    &\expVal{\sum_{n=1}^{N} \expValDist{\min\{C_{a_n}, T_{a_n}\}}{C_{a_n}} \middle| \theta}
    \\
    \nonumber
    &
    =
    \expVal{\sum_{a\in\mathcal{A}} \expValDist{\min\{C_{a}, T_{a}\}}{C_{a}} \times N_a \middle| \theta}
    \\
    \nonumber
    &
    =
    \sum_{a\in\mathcal{A}} \expValDist{\min\{C_{a}, T_{a}\}}{C_{a}}\times\expVal{N_a | \theta}
    \\
    \label{eq: kappa helper 1}
    &
    =
    \sum_{a\in\mathcal{A}} \frac{\expValDist{\min\{C_{a}, T_{a}\}}{C_{a}}}{F_{C_a}(T_a)}\times\expVal{N_{eff}^a | \theta}
    .
\end{align}
Let $\kappa_a(T_a)\triangleq \expValDist{\min\{C_{a}, T_{a}\}}{C_{a}} / F_{C_a}(T_a)$.
We will refer to $\kappa_a$ as the “updated fixed cost” to emphasize that it is a \textit{deterministic user-tunable} value, and that it plays the role of the fixed action cost $c_a$ in Section \ref{section: model and problem review}. 
Notably, regardless of whether $\expVal{C_a} < \infty$ or not for some action $a$, the new objective function is always finite as long as $\expVal{N} < \infty$ since $\kappa_a(T_a)$ is always finite.

The non-canceled action rate is embodied in the denominator of $\kappa_a(T_a)$.
Therefore, studying $\kappa_a(T_a)$ can characterize when action cancellation boosts or degrades performance.
However, the curvature of $\kappa_a$ with respect to $T_a$ is ambiguous.
In fact, depending on the original distribution of $C_a$, the new objective can be either convex, concave, neither, or even completely independent of $T_a$.
In the following lemma, we characterize when the action cancellation does not degrade performance:
\begin{lemma}
    \label{lemma: kappa leq expValC}
    $\kappa_a(T_a) \leq \expVal{C_a}$ if and only if $\expVal{C_a} \leq \expValDist{C_a-T_a|C_a>T_a}{C_a}$.
\end{lemma}
\begin{IEEEproof}
    \ifShowSupp
        See Appendix \ref{lemma: kappa leq expValC proof}.
    \fi   
\end{IEEEproof}
\textit{Remark}: Generally, there are two more (simpler) cases to consider, and Lemma \ref{lemma: kappa leq expValC} only addresses the more challenging case.
The simpler cases are: (i) When $\expVal{C_a} = \infty$.
Here, trivially, $\kappa_a(T_a) < \expVal{C_a}$ for any $T_a$ such that $F_{C_a}(T_a) > 0$.
(ii) When $F_{C_a}(T_a) = 1$, which implies that $\min\{C_a, T_a\} = C_a$ with probability 1, hence $\kappa_a(T_a) = \expVal{C_a}$.
Namely, action cancellation is not leveraged.

The quantity $\expValDist{C_a-T_a | C_a > T_a}{C_a}$ is well-studied in the context of reliability theory (e.g., \cite{Sheldon_etal_2005_reliability_IFR_DFR}), as it characterizes the expected remaining cost provided $T_a$ has already been incurred, i.e., \emph{cost overshoot}.

Counterintuitively, the criterion in Lemma \ref{lemma: kappa leq expValC} does not rely on light-tailed or heavy-tailed cost behavior and instead observes how large the overshoot is from the deadline.
We illustrate this with three examples;
The first two examples are light-tailed distributions: the Erlang and Hyperexponential distributions.
In the former, using action cancellation always degrades performance, whereas in the latter, it always boosts performance.
In the third example, we demonstrate that a cost following a Pareto distribution, i.e., a heavy-tailed distribution, can exhibit either a boost or a degradation, depending on the range from which the deadline is selected.

\begin{proposition}
    \label{proposition: Erlang Example}
    Let $C_a\sim\mathrm{Erlang}(k,\lambda)$ for some $2\leq k\in\mathbb{N}$.
    Then, for any $T_a > 0$, $\expValDist{C_a-T_a|C_a>T_a}{C_a} = \expVal{C_a} - g(T_a)$ for some $g(T_a) > 0$.
    Namely, $\kappa_a(T_a) > \expVal{C_a}$.
\end{proposition}
\begin{IEEEproof}
    \ifShowSupp
        See Appendix \ref{proposition: Erlang Example proof}.
    \fi    
\end{IEEEproof}
\begin{proposition}
    \label{proposition: Hyperexponential Example}
    Assume $C_a$ follows the (two-fold) Hyperexponential distribution, i.e.,
    $C_a\sim\begin{cases}
        \mathrm{Exp}(\alpha_a) & \text{w.p. }p \\
        \mathrm{Exp}(\beta_a) & \text{w.p. }1-p
    \end{cases}$ for some $\alpha_a\neq\beta_a > 0$.
    Then, for any $T_a > 0$, $\expValDist{C_a-T_a|C_a>T_a}{C_a} > \expVal{C_a}$.
\end{proposition}
\begin{IEEEproof}
    \ifShowSupp
        See Appendix \ref{proposition: Hyperexponential Example proof}.
    \fi
\end{IEEEproof}

\begin{proposition}
    \label{proposition: Pareto Example}
    Let $C_a\sim\mathrm{Pareto}(x_{\min, a}, \alpha_a)$ with $\alpha_a > 1$.
    Then, $\kappa_a(T_a) \leq \expVal{C_a}$ if and only if $T_a\geq \alpha_a x_{\min, a}$.
\end{proposition}
\begin{IEEEproof}
    \ifShowSupp
        See Appendix \ref{proposition: Pareto Example proof}.
    \fi
\end{IEEEproof}
Specifically, any $T_a < \alpha_a x_{\min, a}$ degrades performance, whereas any $T_a > \alpha_a x_{\min, a}$ boosts performance.

When action cancellation allows improving the cost expectation, optimizing $\kappa_a$ over $T_a$ can be discussed.
Observe that the asymptotic optimal action distribution in Corollary \ref{corollary: expected information gain per expected cost optimization goal} is not coupled to the per-action deadlines $\{T_a\}_a$.
Accordingly, it is possible to optimize the expected total cost by first minimizing the updated fixed costs, $\{\kappa_a\}_a$ (when they are convex in $\{T_a\}_a$), followed by optimizing the expected information gain per expected cost.
For example, we can compute the optimal per-action deadline for Pareto costs:
\begin{lemma}
    \label{lemma: Pareto cost optimal deadline}
    Assume $C_a\sim\mathrm{Pareto}(x_{\min, a}, \alpha_a)$ with $\alpha_a > 1$.
    Then, 
    (i) $\kappa_a(T_a)$ is convex in $T_a$.
    (ii) $T_a^* = x_{\min,a}\times \tau^*(\alpha_a)$, where $1\leq\tau^*(\alpha_a)$ determined numerically from solving $0 = (\alpha-1)\tau^{\alpha}-\alpha^2\tau^{\alpha-1}+1$.
\end{lemma}

\begin{IEEEproof}
    The proof follows a straightforward computation of $\kappa_a(T_a)$ and deriving it twice with respect to $T_a$ (or $\tau_a = T_a/x_{\min, a}$).
    \ifShowSupp
        See Appendix \ref{lemma: Pareto cost optimal deadline proof} for details.
    \fi
\end{IEEEproof}

\begin{figure} [!htbp]
    \centering
    \includegraphics[scale=0.7]{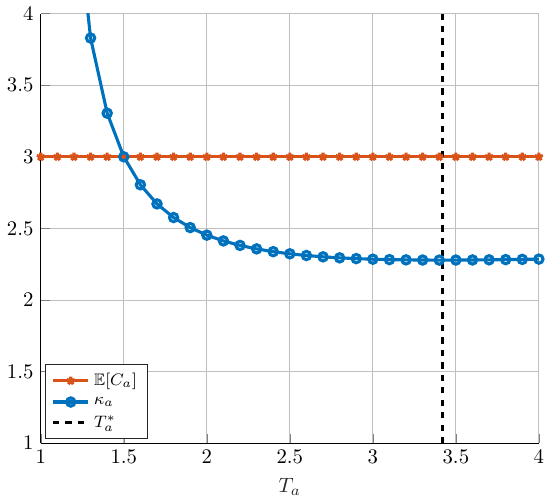}
    \vspace{-8pt}
    \caption{
        Illustrating Proposition \ref{proposition: Pareto Example} when $C_a\sim \mathrm{Pareto}(1, 3/2)$.
        $\kappa_a > \expVal{C_a} = 3$ for any $T_a < 3/2$ and $\kappa_a \leq \expVal{C}$ when $T_a\geq 3/2$.
        The optimal per-action deadline $T_a^* \approx 3.41825$ from Lemma \ref{lemma: Pareto cost optimal deadline}, is also depicted.
    }
    \label{figure: pareto mean vs kappa}
\end{figure}

In Figure \ref{figure: pareto mean vs kappa}, we illustrate Proposition \ref{proposition: Pareto Example} and Lemma \ref{lemma: Pareto cost optimal deadline} by comparing $\kappa_a (T_a)$ with $\expVal{C_a}$ when $C_a\sim \mathrm{Pareto}(1, 3/2)$.
When $T_a < \alpha_a x_{\min, a} = 3/2$, $\kappa_a(T_a) > \expVal{C_a}$, whereas $\kappa_a(T_a) \leq \expVal{C_a}$ when $T_a \geq 3/2$.
It can also be visually verified that $\kappa_a$ is convex in $T_a$, so $T_a^* \approx 3.41825$ (which is the solution for $0.5\tau^{1.5}-2.25\tau^{0.5}+1 = 0$) minimizes $\kappa_a$.

\begin{figure} [!htbp]
    \centering
    \includegraphics[scale=0.7]{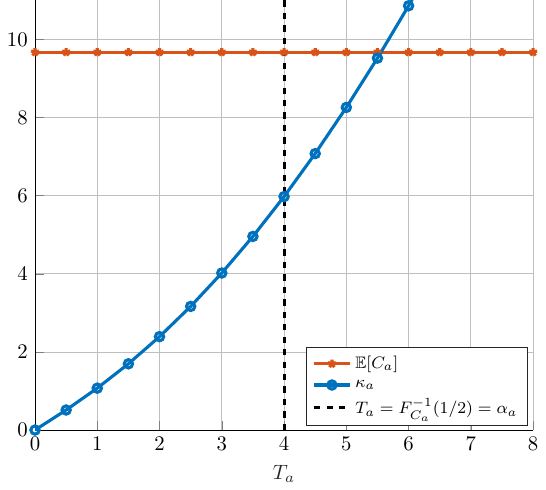}
    \vspace{-8pt}
    \caption{
        Illustrating Proposition \ref{proposition: LogLogistic Example} when $C_a\sim \mathrm{LogLogistic}(4, 3/2)$.
        The updated fixed cost, $\kappa_a$, increases with $T_a$, implying that it cannot be optimized.
        Taking the distribution median $\alpha_a = 4$ as the per-action deadline ensures that $\kappa_a(\alpha_a)\leq\expVal{C_a}$.
    }
    \label{figure: loglogistic mean vs kappa}
\end{figure}

When $\kappa_a$ is not convex, other methodologies should be considered, e.g., taking the per-action deadlines to be the distribution medians as in the Log-Logistic cost models:
\begin{proposition}
    \label{proposition: LogLogistic Example}
    Let $C_a\sim\mathrm{LogLogistic}(\alpha_a,\beta_a)$ with $\beta_a\in(1, 2]$.
    Let $\incompleteBetaFunc{\cdot}{a}{b}$ be the incomplete beta function.
    Then,
    \begin{enumerate}[label=\roman*)]
        \item $\kappa_a(T_a) = (1+(T_a/\alpha_a)^{\beta_a})\frac{\alpha_a}{\beta_a}\incompleteBetaFunc{F_{C_a}(T_a)}{\frac{1}{\beta_a}}{1-\frac{1}{\beta_a}}$
        \item $\kappa_a(T_a)$ strictly increases in $T_a$
        \item If $T_a = \alpha_a$, then $\kappa_a(\alpha_a)\leq \expVal{C_a}$ if and only if $\beta_a\in(1, 2]$
    \end{enumerate} 
\end{proposition}
\begin{IEEEproof}
    The first part follows a straightforward computation of $\kappa_a$.
    For the second part, derive $\kappa_a$ with respect to $T_a$ to find a strictly positive derivative.
    For the last part, substituting $T_a=\alpha_a$ results in $\kappa_a\leq\expVal{C_c}$ being reduced to the inequality $\incompleteBetaFunc{1}{1-\frac{1}{\beta_a}}{1+\frac{1}{\beta_a}} \geq \frac{2}{\beta_a}\incompleteBetaFunc{\frac{1}{2}}{\frac{1}{\beta_a}}{1-\frac{1}{\beta_a}}$ which holds for any $\beta_a\in(1, 2]$.
    \ifShowSupp
        See Appendix \ref{proposition: LogLogistic Example proof} for details.
    \fi
\end{IEEEproof}
Intuitively, Proposition \ref{proposition: LogLogistic Example} asserts the usefulness of action cancellation in terms of how heavy the remaining Log-Logistic tail \emph{beyond the median}; action cancellation helps when the tail is heavy ($1 < \beta < 2$), hurts when the tail is light ($\beta > 2$), and does not hurt when $\beta=2$.
Since $\kappa_a(T_a)$ is strictly increasing in $T_a$, the infimum of $\kappa_a(T_a)$ is achieved in the limit $T_a\to0^+$, where the action yields no samples.
Thus, there is no nontrivial optimal per-action deadline; instead, we use the distribution median $T_a=\alpha_a$ as a canonical choice that guarantees $\kappa_a(T_a)\leq\expVal{C_a}$ when $1 < \beta\leq 2$.
Figure \ref{figure: loglogistic mean vs kappa} illustrates Proposition \ref{proposition: LogLogistic Example} when $C_a\sim \mathrm{LogLogistic}(4, 3/2)$.

\section{Numerical Results}
\label{section: Numerical Results}
We present three simulation scenarios motivated by Internet-mediated remote sensing.
In all scenarios, the \DM seeks to identify one of four possible system states (i.e., $H=4$) by querying one of eight sensors, where each query corresponds to an action (i.e., $\abs{\mathcal{A}}=8$).
The sensor observations are exponentially distributed, with rates depending on both the queried sensor and the true hypothesis.
This model is suitable for situations where sensors track inter-event times, waiting times until a threshold crossing, or the lengths of busy periods.

To isolate the effect of the acquisition-cost distribution, the sample distribution parameters are fixed before running the simulations and are shared across all scenarios.
The complete simulation setup is given in Table \ref{table: simulation setup}.
Here, each row corresponds to an action, and each column corresponds to a hypothesis.
For example, if the \DM applies action 6, then it acquires an exponentially distributed sample whose rate is $1/5$ when $\theta=0$, $1/10$ when $\theta\in\{1,2\}$, or $1/100$ when $\theta=3$.
In addition, the action incurs a random acquisition cost $C_6\sim F(2,1.7)$, where $F$ is chosen according to the considered scenario: Erlang, Pareto, or Log-Logistic.
In the Erlang scenario, the second parameter is interpreted as the mean rather than the rate (e.g., $C_6\sim \mathrm{Erlang}(2, 1/1.7)$ in the previous example).

\begin{table}
    \centering
    \begin{tabular}{|c|c|c|c|c|}
        \hline
         & $H_0$ & $H_1$ & $H_2$ & $H_3$ \\
         \hline
         \makecell{Action 1\\ $C_1\sim F(3, 1.2)$}& Exp$\left(\frac{1}{15}\right)$ & Exp$\left(\frac{1}{15}\right)$ & Exp$\left(\frac{1}{15}\right)$   & Exp$\left(\frac{1}{100}\right)$   \\
         \hline
         \makecell{Action 2\\ $C_2\sim F(2, 1.3)$}& Exp$\left(\frac{1}{30}\right)$ & Exp$\left(\frac{1}{10}\right)$ & Exp$\left(\frac{1}{10}\right)$   & Exp$\left(\frac{1}{100}\right)$   \\
         \hline
         \makecell{Action 3\\ $C_3\sim F(3, 1.5)$}& Exp$\left(\frac{1}{20}\right)$ & Exp$\left(\frac{1}{10}\right)$ & Exp$\left(\frac{1}{10}\right)$   & Exp$\left(\frac{1}{100}\right)$   \\
         \hline
         \makecell{Action 4\\ $C_4\sim F(2, 1.8)$}& Exp$\left(\frac{1}{20}\right)$ & Exp$\left(\frac{1}{20}\right)$ & Exp$\left(\frac{1}{10}\right)$   & Exp$\left(\frac{1}{100}\right)$   \\
         \hline
         \makecell{Action 5\\ $C_5\sim F(3, 1.1)$}& Exp$\left(\frac{1}{15}\right)$ & Exp$\left(\frac{1}{10}\right)$ & Exp$\left(\frac{1}{5}\right)$    & Exp$\left(\frac{1}{100}\right)$   \\
         \hline
         \makecell{Action 6\\ $C_6\sim F(2, 1.7)$}& Exp$\left(\frac{1}{15}\right)$ & Exp$\left(\frac{1}{10}\right)$ & Exp$\left(\frac{1}{10}\right)$   & Exp$\left(\frac{1}{100}\right)$   \\
         \hline
         \makecell{Action 7\\ $C_7\sim F(3, 1.4)$}& Exp$\left(\frac{1}{20}\right)$ & Exp$\left(\frac{1}{15}\right)$ & Exp$\left(\frac{1}{12.5}\right)$ & Exp$\left(\frac{1}{100}\right)$   \\
         \hline
         \makecell{Action 8\\ $C_8\sim F(4, 2.0)$}& Exp$\left(\frac{1}{30}\right)$ & Exp$\left(\frac{1}{20}\right)$ & Exp$\left(\frac{1}{20}\right)$   & Exp$\left(\frac{1}{100}\right)$   \\
         \hline
    \end{tabular}
    \hspace{2pt}
    \caption{Simulations Setup}
    \label{table: simulation setup}
\end{table}

As mentioned before, the three cost models correspond to different remote-sensing setups.
The Erlang scenario models a controller that selects paths in its network to obtain a sensor reading; each sensor resides at a different location, and the total acquisition time is the sum of several exponentially distributed per-hop processing times (e.g., \cite[Chapter~2]{kleinrock1974queueing}).
The Pareto scenario models proxy-based sensing, where some requests require retrieving fresh measurements and may therefore incur heavy-tailed delays \cite{Crovella_Bestavros_1997_ParetoInternetDelay}.
The Log-Logistic scenario models direct querying of remote sensors with heterogeneous end-to-end response times \cite{GagoBenitez_et_al_2013_LogLogisticInternetDelay}.

For each cost model, we compare the classic Chernoff and NJ1 against \CACH and \CANJ, both with and without action cancellation.
When cancellation is enabled, each action is assigned a deadline: for Pareto costs, we use the optimal per-action deadline (evaluated numerically by solving the equation in Lemma \ref{lemma: Pareto cost optimal deadline}), whereas for Erlang and Log-Logistic costs, we use the distribution median acquisition time.
The cost parameters are chosen so that the vanilla and non-canceling \CACH and \CANJ will have a clear edge over the canceling counterpart of the \CA variants in the Erlang settings, highlighting that action cancellation may degrade performance even when the underlying action-selection rule is effective.

\begin{figure}
    \centering
    \includegraphics[scale=0.7]{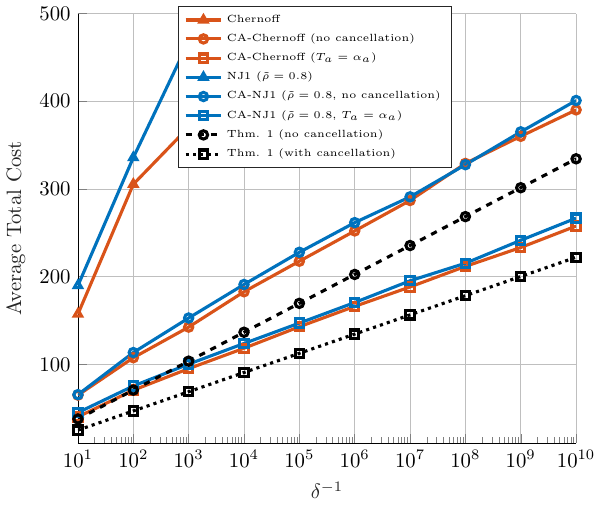}
    \vspace{-8pt}
    \caption{
        Simulation results for the scenario in which each cost follows the Log-Logistic distribution for any action.
        Using the median as the per-action deadline for each action (see Proposition \ref{proposition: LogLogistic Example}) improves performance over the standard \CA algorithms.
        The vanilla algorithms are significantly outperformed as they do not consider cost statistics.
    }
    \label{figure: average total cost loglogistic costs}
\end{figure}

\begin{figure}
    \centering
    \includegraphics[scale=0.7]{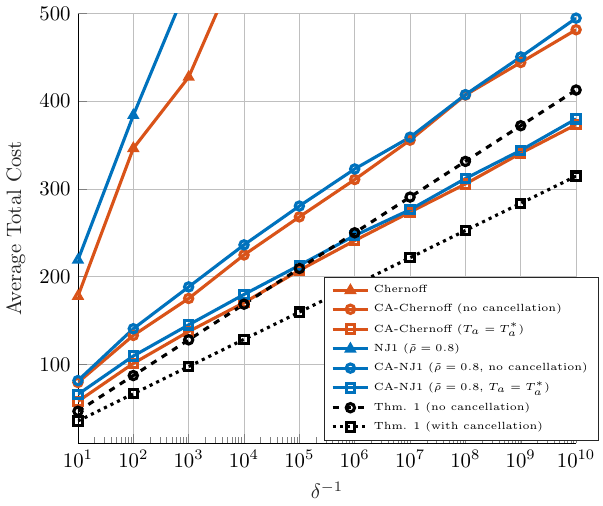}
    \vspace{-8pt}
    \caption{
        Simulation results for the scenario in which the costs follow a Pareto distribution for any action.
        Using the optimal per-action deadline (computed as in Lemma \ref{lemma: Pareto cost optimal deadline}) further improves the performance of \CA algorithms, which outperform their vanilla versions.
    }
    \label{figure: average total cost pareto costs}
\end{figure}

\begin{figure}
    \centering
    \includegraphics[scale=0.7]{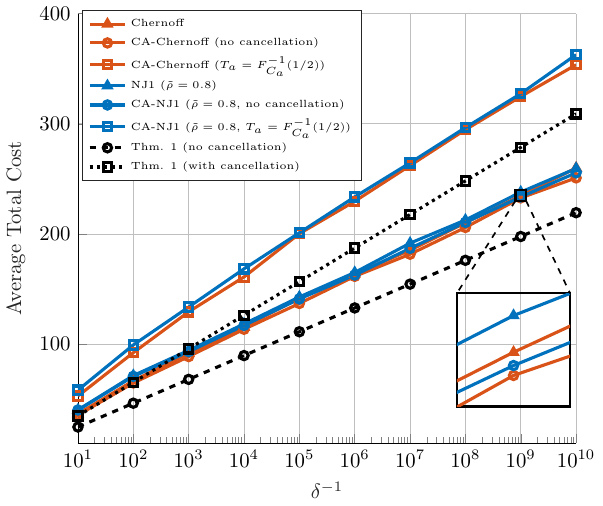}
    \vspace{-8pt}
    \caption{
        Simulation results for the Erlang-distributed-cost scenario for any action.
        Using any per-action deadline degrades performance (see Proposition \ref{proposition: Erlang Example}), highlighting that action cancellation may degrade performance.
    }
    \label{figure: average total cost erlang costs}
\end{figure}

The simulation results are presented in Figures \ref{figure: average total cost loglogistic costs}, \ref{figure: average total cost pareto costs}, and \ref{figure: average total cost erlang costs}.
Across all figures, the orange curves correspond to the Chernoff scheme and its variants, while the blue curves correspond to NJ1 and its variants.
The markers distinguish between the different versions of each algorithm: triangles denote the vanilla schemes, circles denote the corresponding \CA variants, and squares denote the \CA variants with action cancellation.
The black curves represent the oracle bound in Theorem \ref{theorem: expected total cost bounds}, omitting the $1+\littleO{1}$ term, and with optimized action distribution.
Specifically, the curve marked with circles evaluates this expression using the expected action costs, whereas the curve marked with squares uses the updated fixed costs, corresponding to the asymptotic formulations without and with action cancellation, respectively.

The simulation results show that the \CA variants consistently outperform their vanilla counterparts.
Even in the Erlang scenario, where the cost parameters were chosen so that the expected gains from cost awareness are relatively small, the \CA variants reduce the expected total cost by approximately 1.5\%-3\% (see close-up in Figure \ref{figure: average total cost erlang costs}).
The improvement is substantially more pronounced in the Log-Logistic and Pareto scenarios.
In these cases, the expected total costs attained by the vanilla variants, approximately $2450$ and $2280$, are reduced by the corresponding \CA variants to approximately $400$ and $500$, yielding improvements of about 83\% and 78\%, respectively.

The effect of action cancellation is more nuanced.
As suggested by Propositions \ref{proposition: Erlang Example}, \ref{proposition: Pareto Example}, and \ref{proposition: LogLogistic Example}, cancellation can either improve or degrade performance, depending on the distribution of acquisition costs.
This behavior is reflected in the simulations.
In the Erlang scenario, the \CA variants with action cancellation are outperformed by their non-canceling counterparts.
In contrast, in the Log-Logistic and Pareto scenarios, cancellation further reduces the expected total cost from approximately $400$ and $500$ to approximately $266$ and $380$, corresponding to additional improvements of about 33\% and 24\%, respectively.
This conclusion is also reflected in the ordering of the asymptotic benchmarks: in the Log-Logistic and Pareto scenarios, the black curve corresponding to action cancellation lies below the one without cancellation, whereas in the Erlang scenario, the ordering is reversed.

Finally, observe that the x-axis in all figures is logarithmic.
The \CA variants, both with and without action cancellation, exhibit slopes that match their respective asymptotic characterization, supporting their asymptotic optimality and alignment with the expected total cost objective.
By contrast, in the Log-Logistic and Pareto scenarios, the vanilla variants have much steeper slopes, indicating that they are not optimized for the correct objective.
The Erlang scenario is the exception: there, the vanilla and \CA variants have similar slopes because the expected action costs are close in value, making the \CA and cost-oblivious objectives nearly aligned.

\section{Conclusion}
\label{section: conclusion}
In this work, we introduced a cost-aware variant of active \SHT, termed \CASHT, in which each action incurs a positive random acquisition cost that may not reflect the action’s informativeness.
Under this model, we studied the expected total cost of \MSPRT-based procedures, beginning with fixed deterministic costs and then extending the analysis to random costs drawn afresh at each sample acquisition.
In this analysis, we proved that minimizing the expected total cost is equivalent to maximizing the ratio of the expected information gained per sample to the expected cost per sample, evaluated under the action-selection distribution induced by the algorithm (Theorem \ref{theorem: expected total cost bounds}), with Corollary \ref{corollary: expected information gain per expected cost optimization goal} embodying the design principle.
Notably, Theorem \ref{theorem: expected total cost bounds} implies that \MSPRT procedures are essentially optimal in the sense that they achieve optimal scaling, and adapting them to \CASHT entails improving scaling constants.

For random acquisition costs, we considered two cost-revelation models:
The ex-post model, in which the cost is revealed only after the sample is acquired, and the ex-ante model, in which it is revealed before/during acquisition.
The latter enables the \DM to abort an action before receiving its sample and then continue operating as usual.
Such preemption should be used only when it reduces the expected cost of acquiring a successful sample, leading us to derive the mean-overshoot criterion in Lemma \ref{lemma: kappa leq expValC}.

Finally, we adapted two classic active \SHT algorithms, the Chernoff scheme and NJ1, to the \CASHT settings.
The resulting cost-aware algorithms were empirically shown to outperform their classic counterparts while maintaining controllable error probabilities.
Moreover, their observed behavior in simulations agrees with the theoretical characterization of the expected total cost under \MSPRT, demonstrating that the proposed adaptations are aligned with the appropriate cost-aware optimization objective.

\ifShowSupp
    \appendices
\section{Miscellaneous Proofs}
\label{section: misc proofs}

\subsection{Proof of Theorem \ref{lemma: expected total cost decomposition to expected product}}
\label{lemma: expected total cost decomposition to expected product proof}

Since $\expVal{N|\Gamma}<\infty$, we have $\expVal{N|\theta,\Gamma} < \infty$ by the Smoothing Theorem \cite[Section~3.4.2]{leongarcia2008}: $\expVal{N|\Gamma} = \expValDist{ \expValDist{N | \theta,\Gamma}{N} }{\theta}$.
Thus, if each action is applied $N_a$ times, i.e., $\sum_{a\in\mathcal{A}} N_a = N$, then $\expVal{N_a|\theta,\Gamma} < \infty$ for any $a\in\mathcal{A}$.
Hence, the ratio $\expVal{N_a|\theta,\Gamma}/\expVal{N|\theta,\Gamma}$ is well defined ans so is $\myVec{\lambda}_{\theta,\Gamma} \triangleq \{ \expVal{N_a|\theta,\Gamma}/\expVal{N|\theta,\Gamma} \}_{a\in\mathcal{A}}$ for any $\theta\in\mathcal{H}$.
Note that $\myVec{\lambda}_{\theta,\Gamma} \in \probSimplexA$.

Let $A$ be a random variable whose conditional distribution over $\theta$ is $\myVec{\lambda}_{\theta,\Gamma}$, i.e., $A|\theta\sim\myVec{\lambda}_{\theta,\Gamma}$.
Thus,
\begin{align*}
    &\expVal{\sum_{n=1}^{N} c_{A_n} \middle| \theta,\Gamma}
    =
    \expVal{\sum_{a\in\mathcal{A}} c_{a}N_a \middle| \theta,\Gamma}
    \\
    &\quad\quad\quad\quad\quad\quad
    =
    \sum_{a\in\mathcal{A}} c_{a}\times\expVal{N_a | \theta,\Gamma}
    \\
    &\quad\quad\quad\quad\quad\quad
    =
    \expVal{N|\theta,\Gamma} \times \sum_{a\in\mathcal{A}} c_{a} \times\frac{\expVal{N_a | \theta,\Gamma}}{\expVal{N|\theta,\Gamma}}
    \\
    &\quad\quad\quad\quad\quad\quad
    =
    \expVal{N|\theta,\Gamma} \times \sum_{a\in\mathcal{A}} c_{a} \times\vectorComponent{\lambda}{a}[\theta,\Gamma]
    \\
    &\quad\quad\quad\quad\quad\quad
    =
    \expVal{N|\theta,\Gamma} \times \expValDist{c_{A} | \theta,\Gamma}{A\sim\myVec{\lambda}_{\theta, \Gamma}}
    .
\end{align*}
By averaging over $\theta$, we obtain the desired result.

\subsection{Proof of Proposition \ref{proposition: adjusted Chernoff dist}}
\label{proposition: adjusted Chernoff dist proof}
With the new notation, Eq. \eqref{eq: Chernoff Policy Dist With Costs} can be rewritten as
\begin{equation}
    \label{eq: Chernoff Policy Dist With Costs 2}
    \begin{split}
        \max_{ \myVec{\lambda}_i\in\mathbb{R}^{\abs{\mathcal{A}}} }\quad&
            \frac{ \min_{j\neq i} \myVec{d}_{ij}^T\myVec{\lambda}_i }
            { \myVec{c}^T\myVec{\lambda}_i }
        \\
        \mathrm{s. t.}\quad&    
            \begin{cases}
                \myVec{1}^T\myVec{\lambda}_i = 1 \\
                \myVec{\lambda}_i\geq \myVec{0}
            \end{cases}
    \end{split}
    .
\end{equation}
We now show equivalence.
For any $\myVec{\nu}$ feasible in Eq. \eqref{eq: Chernoff Policy Dist With Costs 2}, the pair $\myVec{y} = \myVec{\nu}/(\myVec{c}^T\myVec{\nu})$ and $z = 1/(\myVec{c}^T\myVec{\nu})$ is feasible for the optimization problem in Proposition \ref{proposition: adjusted Chernoff dist}.
The value of the objective function in the problem in Proposition \ref{proposition: adjusted Chernoff dist} becomes $\min_{j\neq i} \myVec{d}_{ij}^T\myVec{\nu} /(\myVec{c}^T\myVec{\nu})$ in this case, which is the same value the objective function in Eq. \eqref{eq: Chernoff Policy Dist With Costs 2} obtains when setting $\myVec{\lambda}_i = \myVec{\nu}$.

Assume $\myVec{y}$ and $z$ are feasible for the problem in Proposition \ref{proposition: adjusted Chernoff dist}.
Note that $z = \myVec{1}^T \myVec{y} \neq 0$ (otherwise, $\myVec{y}=\myVec{0}$ and the constraint $\myVec{c}^T\myVec{y} = 1$ does not hold).
We observe that $\myVec{\nu} = \myVec{y}/z\geq\myVec{0}$ is feasible for Eq. \eqref{eq: Chernoff Policy Dist With Costs 2} as $\myVec{1}^T\myVec{\nu} = \myVec{1}^T \myVec{y}\times\frac{1}{z} = \myVec{1}^T \myVec{y} \times (\myVec{1}^T \myVec{y})^{-1} = 1$.
Particularly, setting $\myVec{\lambda}_i = \myVec{y}/z$ inside the objective function in Eq. \eqref{eq: Chernoff Policy Dist With Costs 2} results in the value of $\min_{j\neq i} \myVec{d}_{ij}^T\myVec{y}$, which is the same value as of the objective function in Proposition \ref{proposition: adjusted Chernoff dist}.


\subsection{Proof of Proposition \ref{proposition: sum Mn = o(n)}}
\label{proposition: sum Mn = o(n) proof}
Observe that $\sum_{t=1}^n M_t$ is a Martingale with respect to $\mathcal{F}_n$, and 
Assumption \ref{assumption: finite LLR variance} allows using the Strong Law of Large Numbers in \cite[Exercise~4.4.11]{durrett2019probability}, which yields the result.

\subsection{Proof of Proposition \ref{proposition: Jj = o(n)}}
\label{proposition: Jj = o(n) proof}
Observe that $I_k (\myVec{\lambda}_{k}^{\mathrm{CA-CH}}) > 0$ implies that $D_{k,k} > 0$.
Specifically, from the minimality of $I_k$, we have $\expValDist{ \KLD{f_{k}^A}{f_\theta^A} }{A\sim \myVec{\lambda}_{k}^{\mathrm{CA-CH}}} > 0$.
Thus, there is non-zero mass on some action $b\in\mathcal{A}$ such that $\KLD{f_{k}^{b}}{f_\theta^{b}} > 0$ (this is also a result of Assumption \ref{assumption: validity}).
Since $\KLD{f}{g} = 0$ if and only if $f=g$ almost everywhere, $\KLD{f_\theta^{b}}{f_{k}^{b}} > 0$.
Since $b$ obtains non-zero mass, $D_{k,k} > 0$.

Let $\tau_k\triangleq\sup\{n\in\mathbb{N} : i_n^* = k\} = J_k(\infty)$ be the last timestep in which hypothesis $k$ guides the action distribution.
At $\tau_k$, the accumulated \LLR between $\theta$ and $k$ is negative, i.e., $0\geq \sum_{t=1}^{\tau_k} M_t + D_{k,k} J_k(\tau_k)$.
Thus, $D_{k,k} J_k(\tau_k) \leq -\sum_{t=1}^{\tau_k} M_t \leq \sup_{s\leq n} \abs{\sum_{t=1}^{s} M_t}$, where the latter is $\bigO{\sqrt{n \log \log n}} = \littleO{n}$ from the Law of the Iterated Logarithm \cite[Theorem~8.5.2]{durrett2019probability}.
        
\subsection{Proof of Lemma \ref{lemma: E[W(N)] is bounded}}
\label{lemma: E[W(N)] is bounded proof}
We first disregard the stopping rule and observe that $\mathcal{W}(\infty) = \sum_{j\neq\theta} J_j(\infty)$ is independent of $\delta$ as $\delta$ interacts with the algorithm only in its stopping rule.    
Define $N_0 \triangleq \sup\{n\in\mathbb{N} : i_n^*\neq \theta\} = \sup\{n\in\mathbb{N} : \min_{k\neq\theta} L_{\theta, k} \leq 0\}$ as the timestep from which \CACH always draws actions according to $\myVec{\lambda}_{\theta}^{\mathrm{CA-CH}}$.
Note that $N_0 < \infty$ almost surely since $\min_{k\neq\theta} L_{\theta, k}\to\infty$ almost surely.
Next, note that there can be at most $N_0$ timesteps during which \CACH is misguided.
Consequently, from monotonicity, we have $\mathcal{W}(N) \leq \mathcal{W}(\infty) \leq N_0$.
We now show that $\expVal{N_0} < \infty$ regardless of $\delta$.
Compute
\begin{align*}
    \expVal{N_0}
    &=
    \sum_{n=1}^{\infty} \prob{ N_0 \geq n }
    =
    \sum_{n=1}^{\infty} \prob{ \exists m\geq n : i_m^* \neq \theta }
    \\
    &
    \leq
    \sum_{n=1}^{\infty} \sum_{k\neq\theta} \prob{ \exists m\geq n : L_{\theta, k}\leq 0 }
    ,
\end{align*}
with the last inequality following the union bound and $\{i_m^* \neq \theta\} = \{ \min_{k\neq\theta} L_{\theta, k} \leq 0 \} \subseteq \{ \exists k : L_{\theta, k} \leq 0 \}$.
The following bounds $\prob{ \exists m\geq n : L_{\theta, k}\leq 0 }$ exponentially.
\begin{proposition}
    $\prob{ \exists m\geq n : L_{\theta, k}\leq 0 } \leq \xi \times 2^{-\zeta n}$ for some constants $\xi, \zeta>0$.
\end{proposition}
\begin{IEEEproof}
    Denote by
    \begin{align*}
        \psi_{a, k}(\eta)
        &\triangleq
        -\log \expValDist{2^{-\eta \log \frac{f_\theta^A(X)}{f_k^A(X)}} \middle| A = a }{X\sim f_\theta^a}
        \\
        &=
        -\log \expValDist{\left( \frac{f_k^a(X)}{f_\theta^a(X)} \right)^{\eta}}{X\sim f_\theta^a}
        \\
        &=
        -\log \int_{-\infty}^{\infty} (f_k^a(x))^{\eta} (f_\theta^a(x))^{1-\eta} dx
    \end{align*}
    the (unoptimized) Chernoff exponent \cite[Eq.~(11.239)]{CoverThomas2006}.
    Note that it is non-negative for any $\eta\in(0, 1)$ from Holder’s inequality: $\int_{-\infty}^{\infty} (f_k^a(x))^{\eta} (f_\theta^a(x))^{1-\eta} dx \leq (\int_{-\infty}^{\infty} f_k^a(x) dx)^{\eta} \times (\int_{-\infty}^{\infty} f_\theta^a(x) dx)^{1-\eta} \leq 1$ for any $a$, and is strictly positive if and only if $f_k^a(x) \neq f_\theta^a(x)$.
    Define the tilted \LLR as
    \begin{align*}
        \log \Lambda_m^{\theta, k}(\eta) \triangleq
        -\eta L_{\theta, k}(\indexedSet{X}{1}{m}, \indexedSet{A}{1}{m}) + \sum_{t=1}^m \chi_{t, k}(\eta)
        ,
    \end{align*}
    where $\chi_{t, k}(\eta)$ is the action-tilted Chernoff exponent:
    \begin{align*}
        &\chi_{t, k}(\eta)
        \triangleq
        -\log \expValDist{2^{-\eta \log \frac{f_\theta^{A_t}(X_t)}{f_k^{A_t}(X_t)}} \middle| \mathcal{F}_{t-1} }{A_t, X_t}
    \end{align*}
    that motivates $\phi_j(\eta, \myVec{\lambda}) \triangleq -\log \expValDist{2^{-\psi_{A, j}(\eta)}}{A\sim\myVec{\lambda} }$, which satisfies the following.
    \begin{claim}
        \label{claim: phi j > 0}
        $\phi_j(\eta, \myVec{\lambda}_{\theta}^{\mathrm{CA-CH}}), \phi_j(\eta, \myVec{\lambda}_{j}^{\mathrm{CA-CH}}) > 0$ for any $j\neq\theta$.
    \end{claim}
    \begin{IEEEproof}
        Let $\Delta_{j, \theta}^a(\eta) \triangleq \int_{-\infty}^{\infty} (f_k^a(x))^{\eta} (f_\theta^a(x))^{1-\eta} dx$.
        As an integral of a non-negative function, we have $\Delta_{j, \theta}^a(\eta)\geq 0$.
        From AM-GM inequality, $\Delta_{j, \theta}^a(\eta) \leq \int_{-\infty}^{\infty} \eta f_k^a(x) + (1-\eta) f_\theta^a(x) dx = 1$.
        Namely, $\Delta_{j, \theta}^a(\eta)\in [0, 1]$ for any $\eta\in(0, 1)$, with $\Delta_{j, \theta}^a(\eta)=1$ if and only if $f_j^a = f_\theta^a$.
        
        Suppose that $\phi_j(\eta, \myVec{\lambda}_{j}^{\mathrm{CA-CH}}) = 0$, i.e.,
        \begin{align*}
            1
            &=
            2^{-\phi_j(\eta, \myVec{\lambda}_{j}^{\mathrm{CA-CH}})}
            =
            \expValDist{2^{-\psi_{A, j}(\eta)}}{A\sim\myVec{\lambda}_{j}^{\mathrm{CA-CH}} }
            \\
            &=
            \expValDist{\Delta_{j, \theta}^A(\eta)}{A\sim\myVec{\lambda}_{j}^{\mathrm{CA-CH}}}
            .
        \end{align*}
        The expectation of a random variable taking values on $[0, 1]$ can be $1$ only if the random variable takes the value 1 almost surely.
        Thus, $\myVec{\lambda}_{j}^{\mathrm{CA-CH}}$ assigns probability mass of 1 to some action $b\in\mathcal{A}$ with $f_j^b = f_\theta^b$ and cannot discriminate between $j$ and $\theta$.
        Thus, $I_j (\myVec{\lambda}_{j}^{\mathrm{CA-CH}}) = 0$ - contradiction.
        Showing that $\phi_j(\eta, \myVec{\lambda}_{\theta}^{\mathrm{CA-CH}}) > 0$ is similar.
    \end{IEEEproof}

    Note that $\Lambda_m^{\theta, k}(\eta)$ is a Martingale with with respect $\mathcal{F}_m$:
    \begin{align*}
        \expVal{\Lambda_m^{\theta, k}(\eta) | \mathcal{F}_{m-1}}
        &=
        \Lambda_{m-1}^{\theta, k}(\eta)
        \times
        \expVal{\left( \frac{f_k^{A_m}(X_m)}{f_\theta^{A_m}(X_m)} \right)^{\eta}}
        \\
        &\quad\quad
        \times
        2^{\psi_m(\eta)}
        =
        \Lambda_{m-1}^{\theta, k}(\eta)
        ,
    \end{align*}
    and that $\expVal{\Lambda_m^{\theta, k}(\eta)} = 1$ regardless of how $A_m$ is chosen.
    Since $\phi_k(\eta, \myVec{\lambda}_{\theta}^{\mathrm{CA-CH}}) > 0$ from Claim \ref{claim: phi j > 0},
    \begin{align}
        \label{eq: CACH EN0 helper 0}
        \sum_{t=1}^m \chi_{t, k}(\eta)
        \geq
            \phi_k(\eta, \myVec{\lambda}_{\theta}^{\mathrm{CA-CH}}) \times (m-\mathcal{W}(m))
            %
        .
    \end{align}
    \textit{Remark}: According to Proposition \ref{proposition: Jj = o(n)}, the last inequality implies that $\log \Lambda_m^{\theta, k}(\eta) \geq m\times w\times(1+o(1))$ for some $w>0$, establishing that the tilted \LLR grows linearly just like the untilted \LLR, which in turn implies that the tilted likelihood, $\Lambda_m^{\theta, k}(\eta)$, grows exponentially.
    This observation will be captured via exponential concentration.
    
    Fix $\epsilon > 0$ and decompose:
    \begin{align*}
        &\prob{ \exists m\geq n : L_{\theta, k}\leq 0 }
        \\
        &\quad
        =
        \prob{ \exists m\geq n : L_{\theta, k}\leq 0, \mathcal{W}(m) \leq \epsilon m }
        \\
        &\quad\quad\quad\quad+
        \prob{ \exists m\geq n : L_{\theta, k}\leq 0, \mathcal{W}(m) > \epsilon m }
        \\
        &\quad
        \leq
        \underbrace{ \prob{ \exists m\geq n : L_{\theta, k}\leq 0, \mathcal{W}(m) \leq \epsilon m }}_{\triangleq E_1}
        \\
        &\quad\quad\quad\quad
        +
        \underbrace{ \prob{ \exists m\geq n : \mathcal{W} > \epsilon m } }_{\triangleq E_2}
        .
    \end{align*}
    Note that $\{L_{\theta, k}\leq 0\} = \{-\eta L_{\theta, k} + \sum_{t=1}^m \chi_{t, k}(\eta)\geq \sum_{t=1}^m \chi_{t, k}(\eta)\} = \{ \log \Lambda_m^{\theta, k}(\eta) \geq \sum_{t=1}^m \chi_{t, k}(\eta) \}$.
    
    We first bound $E_1$.
    On the event $\{\mathcal{W}(m) \leq \epsilon m\}$, we have $\{m-\mathcal{W}(m) \geq (1-\epsilon) m\}$.
    Thus, 
    \begin{align}
        E_1
        \label{eq: CACH EN0 helper 1}
        &\leq
        \prob{ \exists m\geq n :\log \Lambda_m^{\theta, k}(\eta) \geq \phi_k(\eta, \myVec{\lambda}_{\theta}^{\mathrm{CA-CH}})(1-\epsilon)m }
        \\
        \label{eq: CACH EN0 helper 2}
        &\leq
        \prob{ \exists m\geq n :\log \Lambda_m^{\theta, k}(\eta) \geq \phi_k(\eta, \myVec{\lambda}_{\theta}^{\mathrm{CA-CH}})(1-\epsilon)n }
        \\
        \nonumber
        &\leq
        \prob{ \exists m\geq n :\Lambda_m^{\theta, k}(\eta) \geq 2^{\phi_k(\eta, \myVec{\lambda}_{\theta}^{\mathrm{CA-CH}})(1-\epsilon)n} }
        \\
        \label{eq: CACH EN0 helper 3}
        &\leq
        2^{-\phi_k(\eta, \myVec{\lambda}_{\theta}^{\mathrm{CA-CH}})(1-\epsilon)n}
        \leq
        2^{-\zeta_k(1-\epsilon)n}
        ,
    \end{align}
    where Eq. \eqref{eq: CACH EN0 helper 1} follows from combining Eq. \eqref{eq: CACH EN0 helper 0} with the event $\{m-\mathcal{W}(m) \geq (1-\epsilon) m\}$, Eq. \eqref{eq: CACH EN0 helper 2} follows the monotonicity of \CDFs, and the inequalities in Eq. \eqref{eq: CACH EN0 helper 3} used the Markov inequality followed by optimization of $\phi_k$ over $\eta$.

    To bound $E_2$, observe that $\mathcal{W} > \epsilon m$ implies that at least one $j\neq\theta$ has $J_j \geq \epsilon m/(H-1)$ (otherwise, $\mathcal{W}\leq \epsilon m$).
    
    In every timestep in which $i_n^* = j$, the accumulated \LLR between $\theta$ and $j$ is negative, i.e., $\{i_n^* = j\}\subseteq \{ L_{\theta, j}\leq 0 \}$.
    Thus, in these steps, their respective accumulated tilted \LLR, $\log \Lambda_{m}^{\theta, j}$, has a positive drift $\chi_{t, j}(\eta) \geq \phi_j(\eta, \myVec{\lambda}_{j}^{\mathrm{CA-CH}}) > 0$ from Claim \ref{claim: phi j > 0}.
    Summing them yields 
    \begin{align*}
        \sum_{t=1}^m \chi_{t, j}(\eta)
        \geq
        \phi_j(\eta, \myVec{\lambda}_{j}^{\mathrm{CA-CH}}) \times J_j(m)
        .
    \end{align*}
    Consequently, by the union bound over $j\neq\theta$, we have
    \begin{align}
        \nonumber
        &E_2
        \leq
        \sum_{j\neq\theta}\prob{ \exists m\geq n : J_j(m) \geq \frac{\epsilon}{H-1} m }
        \\
        \nonumber
        &\leq
        \sum_{j\neq\theta}
        \mathbb{P}\bigg(
            \exists m\geq n : \log \Lambda_{m}^{\theta, j} \geq \phi_j(\eta, \myVec{\lambda}_{j}^{\mathrm{CA-CH}}) \times J_j(m)
            \\\nonumber
            &\quad\quad\quad\quad\quad\quad\quad\quad\quad\quad\quad\quad\quad\quad
            ,J_j(m) \geq \frac{\epsilon}{H-1} m
        \bigg)
        \\
        \nonumber
        &\leq
        \sum_{j\neq\theta}\prob{ \log \Lambda_{m}^{\theta, j} \geq \phi_j(\eta, \myVec{\lambda}_{j}^{\mathrm{CA-CH}}) \times \frac{\epsilon}{H-1}m }
        \\
        \nonumber
        &\leq
        \sum_{j\neq\theta}\prob{ \log \Lambda_{m}^{\theta, j} \geq \phi_j(\eta, \myVec{\lambda}_{j}^{\mathrm{CA-CH}}) \times \frac{\epsilon}{H-1}n }
        \\
        \nonumber
        &\leq
        \sum_{j\neq\theta} 2^{-\zeta_j^{\prime}\frac{\epsilon}{H-1}n}
        ,
    \end{align}
    The result now follows since $\prob{ \exists m\geq n : L_{\theta, k}\leq 0 }\leq E_1 + E_2 \leq 2^{-\zeta_k(1-\epsilon)n} + \sum_{j\neq\theta} 2^{-\zeta_j^{\prime}\frac{\epsilon}{H-1}n}$.        
\end{IEEEproof}
Since $\prob{ \exists m\geq n : L_{\theta, k}\leq 0 }\leq \xi \times 2^{-\zeta n}$ for some $\xi, \zeta>0$, $\expVal{N_0}\leq \sum_{n=1}^{\infty} (H-1) \times\xi \times 2^{-\zeta n} < \infty$, which in turn implies that $\expVal{\mathcal{W}(N)} < \infty$.
By taking $\iota = \expVal{N_0}$, we obtain the desired result.

\subsection{Proof of Proposition \ref{proposition: adjusted NJ1 exploration dist}}
\label{proposition: adjusted NJ1 exploration dist proof}
Similar to the proof of Proposition \ref{proposition: adjusted Chernoff dist}, the optimization in Eq. \eqref{eq: NJ1 Exploration Dist With Costs} can be re-written as:
\begin{equation}
    \label{eq: NJ1 Exploration Dist With Costs 2}
    \begin{split}
        \max_{ \myVec{\lambda}\in\mathbb{R}^{\abs{\mathcal{A}}} }\quad&
            \min_{i\in\mathcal{H}}
            \frac{ \min_{j\neq i} \myVec{d}_{ij}^T\myVec{\lambda} }
            { \myVec{c}^T\myVec{\lambda} }
        \\
        \mathrm{s. t.}\quad&    
            \begin{cases}
                \myVec{1}^T\myVec{\lambda} = 1 \\
                \myVec{\lambda}\geq \myVec{0}
            \end{cases}
    \end{split}
    .
\end{equation}
Now, checking common feasible points for Eq. \eqref{eq: NJ1 Exploration Dist With Costs 2} and Proposition \ref{proposition: adjusted NJ1 exploration dist} is the same as before; any feasible $\myVec{\nu}$ can be used to construct $\myVec{y} = \myVec{\nu}/(\myVec{c}^T\myVec{\nu})$ and $z=1/(\myVec{c}^T\myVec{\nu})$, which are feasible for Proposition \ref{proposition: adjusted NJ1 exploration dist}.
For the other direction, for any feasible $\myVec{y}$ and $z$ for the optimization problem in Proposition \ref{proposition: adjusted NJ1 exploration dist}, we construct $\nu = \myVec{y}/z$, which is feasible and well-defined as before.
Notably, the objective function values are the same in both cases.

\subsection{Proof of Proposition \ref{proposition: Na = f(Na_eff)}}
\label{proposition: Na = f(Na_eff) proof}
Let $\indicator{A}$ be the indicator function of the event $A$.
Let $\{n_k\}_{k=1}^{N_a}$ denote the indices for which $A_n = a$.
Thus, $N_{eff}^a|\theta = \sum_{k=1}^{N_a|\theta} \indicator{C_{A_{n_k}}\leq T_a}$.
Since this is a sum of i.i.d. Bernoulli random variables independent of $N_a|\theta$, Wald’s Identity yields $\expVal{N_{eff}^a|\theta} = \expVal{N_a|\theta} \times F_{C_a}(T_a)$.

\subsection{Proof of Lemma \ref{lemma: kappa leq expValC}}
\label{lemma: kappa leq expValC proof}
For notational simplicity, the action subscript $a$ is dropped throughout the proof.
Since $\min\{C,T\} = C - (C-T)^+$, where $x^+\triangleq \max\{x, 0\}$, we have
\begin{align*}
    \kappa(T) \leq \mu
    &\iff
    \mu - \expValDist{(C-T)^+}{C} \leq \mu F_C(T)
    \\
    &\iff
    \mu(1-F_C(T)) \leq \expValDist{(C-T)^+}{C}
    .
\end{align*}
From the Smoothing Theorem \cite[Section~3.4.2]{leongarcia2008}, 
\begin{align*}
    \expValDist{(C-T)^+}{C}
    &=
    \expValDist{C-T | C > T}{C} (1-F_C(T))
    \\
    &\quad + \expValDist{0|C\leq T}{C} F_C(T)
    \\
    &=
    \expValDist{C-T | C > T}{C} (1-F_C(T)).
\end{align*}
Thus, $\kappa(T)\leq \mu \iff \mu\leq \expValDist{C-T|C>T}{C}$.

\subsection{Proof of Proposition \ref{proposition: Erlang Example}}
\label{proposition: Erlang Example proof}

Assume $2\leq k\in\mathbb{N}$.
We drop the subscript $a$ for notational simplicity.
Recall the complementary \CDF of the Erlang distribution: $\prob{C>T} = 1-F_{C}(T) = e^{-\lambda T}\sum_{n=0}^{k-1}\frac{(\lambda T)^n}{n!}$.
Recall its \PDF: $f_C(t) = f_{C}(t;k,\lambda) = \frac{\lambda^kt^{k-1}}{(k-1)!}e^{-\lambda T}$.
Recall the the expected cost is $\expVal{C} = \frac{k}{\lambda}$.

We now compute $\expVal{C|C>T} = \frac{\expValDist{C\times\indicator{C>T}}{C}}{\prob{C>T}}$.
Observe that $t\times f_{C}(t;k,\lambda) = t\times\frac{\lambda^kt^{k-1}}{(k-1)!}e^{-\lambda T} = \frac{k}{\lambda} f_{C}(t;k+1,\lambda)$.
Thus, $\expValDist{C\times\indicator{C>T}}{C} = \int_{T}^\infty t f_{C}(t;k,\lambda) dt = \frac{k}{\lambda}\times(1-F_{C}(T;k+1,\lambda)) = \frac{k}{\lambda}\times e^{-\lambda T}\sum_{n=0}^{k}\frac{(\lambda T)^n}{n!}$.
Accordingly:
\begin{align*}
    \expValDist{C-T|C>T}{C}
    &=
    \frac{k}{\lambda}
    \times
    \frac{\sum_{n=0}^{k}\frac{(\lambda T)^n}{n!}}{\sum_{n=0}^{k-1}\frac{(\lambda T)^n}{n!}}-T
    \\
    &=
    \frac{k}{\lambda}
    \times
    \left(
        1+\frac{\frac{(\lambda T)^k}{k!}}{\sum_{n=0}^{k-1}\frac{(\lambda T)^n}{n!}}
    \right) -T
    \\
    &=
    \frac{k}{\lambda} + \frac{ \frac{(\lambda T)^k}{(k-1)!} }{\lambda\sum_{n=0}^{k-1}\frac{(\lambda T)^n}{n!}} -T
    \\
    &=
    \frac{k}{\lambda} + 
    \frac{ \frac{(\lambda T)^k}{(k-1)!} - \lambda T \sum_{n=0}^{k-1}\frac{(\lambda T)^n}{n!} }{\lambda\sum_{n=0}^{k-1}\frac{(\lambda T)^n}{n!}}
    \\
    &=
    \frac{k}{\lambda} - 
    \frac{ \sum_{n=0}^{k-2}\frac{(\lambda T)^{n+1}}{n!} }{\lambda\sum_{n=0}^{k-1}\frac{(\lambda T)^n}{n!}}
    <
    \frac{k}{\lambda}
    =
    \expVal{C}
    .
\end{align*}

In Figure \ref{figure: erlang mean vs conditional mean}, we visualize the gap $\expValDist{C-T|C>T}{C} - \expVal{C}$ when $C\sim\mathrm{Erlang}(2, 1)$ for $T\in[0, 2]$.
The curves coincide only at $T=0$, but selecting this value yields no samples.

\begin{figure} [!htbp]
    \centering
    \includegraphics[scale=0.7]{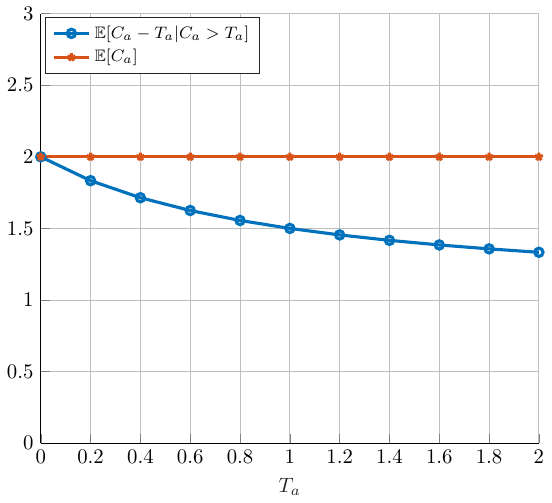}
    \vspace{-8pt}
    \caption{
        $\expValDist{C_a-T_a|C_a>T_a}{C_a}$ and $\expVal{C_a}$ when $C_a\sim\mathrm{Erlang}(2, 1)$.
        For any $T_a>0$, since $\expValDist{C_a-T_a|C_a>T_a}{C_a} < \expVal{C_a}$, $\kappa(T_a) > \expVal{C_a}$.
    }
    \label{figure: erlang mean vs conditional mean}
\end{figure}

\subsection{Proof of Proposition \ref{proposition: Hyperexponential Example}}
\label{proposition: Hyperexponential Example proof}
We drop the subscript $a$ for notational simplicity.
We first show that
$C\sim
\begin{cases}
    \mathrm{Exp}(\alpha) & \text{w.p. }p \\
    \mathrm{Exp}(\beta) & \text{w.p. }1-p
\end{cases}$ is a light-tailed distribution.
The cost’s complementary \CDF at $t$ is $1-F_C(t) = p e^{-\alpha t} + (1-p)e^{-\beta t}$.
Thus, the limit
\begin{align*}
    \lim_{t\to\infty} e^{s t} (1-F_C(t))
    &=
    \lim_{t\to\infty} e^{s t} (p e^{-\alpha t} + (1-p)e^{-\beta t})
    \\
    &=
    \lim_{t\to\infty} p e^{(s-\alpha) t} + (1-p)e^{(s-\beta) t}
    \\
    &< \infty
\end{align*}
for any $s \leq \min\{\alpha, \beta\}$.
Hence, $C$ is light-tailed.

The mean cost is $\expVal{C} = p\times\frac{1}{\alpha} + (1-p)\times\frac{1}{\beta}$.
Compute:
\begin{align*}
    \expValDist{C-T|C>T}{C}
    &=
    \frac{\expValDist{(C-T)^+}{C}}{1-F_C(T)}
    \\
    &=
    \frac{ \int_T^\infty (1-F_C(t)) dt }{ 1-F_C(T) }
    \\
    &=
    \frac{ \int_T^\infty (p e^{-\alpha t} + (1-p)e^{-\beta t}) dt }{ p e^{-\alpha T} + (1-p)e^{-\beta T} }    
    \\
    &=
    \frac{ p e^{-\alpha T}\times\frac{1}{\alpha} + (1-p)e^{-\beta T}\times\frac{1}{\beta} }{ p e^{-\alpha T} + (1-p)e^{-\beta T} }
    \\
    &=
    \eta (T) \times \frac{1}{\alpha} + (1-\eta(T))\times \frac{1}{\beta}
\end{align*}
where $\eta(T) \triangleq \frac{p e^{-\alpha T}}{p e^{-\alpha T} + (1-p)e^{-\beta T}} = \frac{1}{1+\frac{1-p}{p}e^{(\alpha-\beta)T} }$.
Observe the difference $\expValDist{C-T|C>T}{C} - \expVal{C}$:
\begin{align*}
    &(\eta (T)-p) \times \frac{1}{\alpha} + (1-\eta(T)-1+p)\times \frac{1}{\beta}
    \\
    &\quad\quad
    =
    (\eta (T)-p) \times \frac{1}{\alpha} - (\eta(T)-p)\times \frac{1}{\beta}
    \\
    &\quad\quad
    =
    (\eta (T)-p) \times \left(\frac{1}{\alpha} - \frac{1}{\beta}\right)
    .
\end{align*}
If $\alpha < \beta$, then $\frac{1}{\alpha} - \frac{1}{\beta} > 0$ and $e^{-\alpha T}>e^{-\beta T}$, which implies:
\begin{align*}
    \eta(T) - p
    &=
    \frac{p e^{-\alpha T}}{p e^{-\alpha T} + (1-p)e^{-\beta T}} - p
    \\
    &=
    \frac{p e^{-\alpha T} -p^2e^{-\alpha T}-(1-p)pe^{-\beta T}}{p e^{-\alpha T} + (1-p)e^{-\beta T}}
    \\
    &=
    \frac{(1-p)p( e^{-\alpha T}-e^{-\beta T})}{p e^{-\alpha T} + (1-p)e^{-\beta T}} > 0,
\end{align*}
i.e., $\expValDist{C-T|C>T}{C} > \expVal{C}$ for any $T>0$.
Similarly, if $\alpha > \beta$, then $\frac{1}{\alpha} - \frac{1}{\beta} < 0$ and $\eta(T) - p < 0$.
That is, $\expValDist{C-T|C>T}{C} > \expVal{C}$ for any $T>0$ once again.

In Figure \ref{figure: hyperexp mean vs conditional mean}, we show $\expValDist{C-T|C>T}{C}$ as a function of $T$ when $\alpha=2$, $\beta=3$, and $p=0.7$.

\begin{figure} [!htbp]
    \centering
    \includegraphics[scale=0.7]{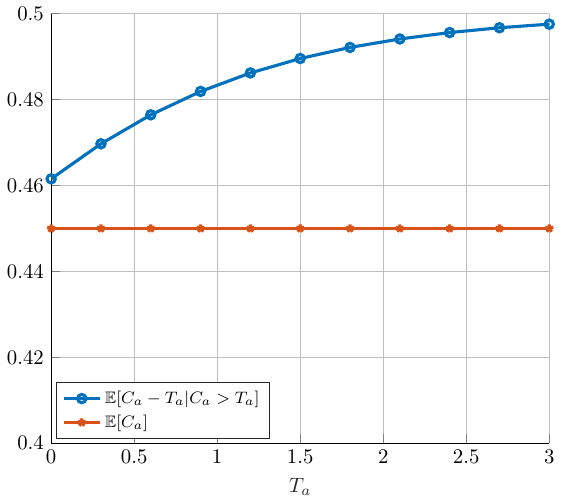}
    \vspace{-8pt}
    \caption{
        $\expValDist{C_a-T_a|C_a>T_a}{C_a}$ and $\expVal{C_a}$ when $C_a$ follows the Hyperexponential distribution (with probability $p=0.7$, $C_a\sim\mathrm{Exp}(2)$, and with probability $1-p=0.3$, $C_a\sim\mathrm{Exp}(3)$).
        Since $\expValDist{C_a-T_a|C_a>T_a}{C_a} > \expVal{C_a}$ for any $T_a>0$, it follows that $\kappa(T_a) < \expVal{C_a}$.
    }
    \label{figure: hyperexp mean vs conditional mean}
\end{figure}

\subsection{Proof of Proposition \ref{proposition: Pareto Example}}
\label{proposition: Pareto Example proof}
We again drop the subscript $a$ for notational simplicity.
Assume $\alpha > 1$.
Since we assume that $F_C(T) > 0$, we have $T > x_{\min}$.
Recall the complementary \CDF of the Pareto distribution: $1-F_C(T) = (x_{\min}/T)^{\alpha}$.
Recall that $\expVal{C} = \frac{\alpha}{\alpha-1}x_{\min}$.
Compute:
\begin{align*}
    \prob{ C > c | C > T}
    =
    \begin{cases}
        \frac{\prob{C>c}}{\prob{C>T}}
        =
        \left(\frac{T}{c}\right)^\alpha
        & c\geq T
        \\
        0 & c < T
    \end{cases}
    .
\end{align*}
Thus, $C|C>T\sim\mathrm{Pareto}(T, \alpha)$.
Hence, its mean is $\frac{\alpha}{\alpha-1}T$.
Accordingly, $\expValDist{C-T | C>T}{C} = \frac{\alpha}{\alpha-1}T - T = \frac{1}{\alpha-1}T$.
Leveraging Lemma \ref{lemma: kappa leq expValC}, $\kappa(T)\leq \expVal{C}$ if and only if $\frac{1}{\alpha-1}T \geq \frac{\alpha}{\alpha-1}x_{\min}$, i.e., when $T\geq \alpha x_{\min}$.

\subsection{Proof of Lemma \ref{lemma: Pareto cost optimal deadline}}
\label{lemma: Pareto cost optimal deadline proof}
We drop the subscript $a$ for notational simplicity.
Following the decomposition $\expValDist{\min\{C,T\}}{C} = \expVal{C} - \expValDist{(C-T)^+}{C} = \expVal{C} - \expValDist{C-T|C>T}{C}(1-F_C(T))$ and our calculation in \ref{proposition: Pareto Example proof}, the denominator for $\kappa(T)$ becomes
\begin{align*}
    \frac{\alpha x_{\min}}{\alpha-1} - \frac{T}{\alpha-1}\times\left(\frac{x_{\min}}{T}\right)^{\alpha}
    =
    \frac{x_{\min}}{\alpha-1}\left(\alpha - \left(\frac{x_{\min}}{T}\right)^{\alpha-1}\right)
\end{align*}
Let $\tau \triangleq T/x_{\min}$.
With this definition, $F_C(T)$ becomes $1-\tau^{-\alpha}$, and $\tau\geq1$ (i.e., $\tau^{\alpha}\geq1$).
Thus,
\begin{align*}
    \kappa(\tau)
    =
    \frac{x_{\min}}{\alpha-1}\times\frac{ \alpha - \tau^{1-\alpha} }{ 1-\tau^{-\alpha} }
    =
    \frac{x_{\min}}{\alpha-1}\times\frac{ \alpha\tau^{\alpha} - \tau }{ \tau^{\alpha}-1 }
    .
\end{align*}
For the rest of the proof, we will ignore the (positive) scalar $\frac{x_{\min}}{\alpha-1}$.
Deriving with respect to $\tau$:
\begin{align*}
    \partialDerive{\tau}\left(\frac{ \alpha\tau^{\alpha} - \tau }{ \tau^{\alpha}-1 }\right)
    =
    \frac{ (\alpha-1)\tau^{\alpha}-\alpha^2\tau^{\alpha-1}+1 }{(\tau^{\alpha}-1)^2}
    =
    0
    .
\end{align*}
Accordingly, $\tau^*$ is a solution of $(\alpha-1)\tau^{\alpha}-\alpha^2\tau^{\alpha-1}+1$.
For convexity, we derive again:
\begin{align*}
    \partialDeriveSecond{\tau}\left(\frac{ \alpha\tau^{\alpha} - \tau }{ \tau^{\alpha}-1 }\right)
    &=
    \frac{\alpha\tau^{\alpha-2}}{(\tau^{\alpha}-1)^3}
    \times
    (
        \alpha (\tau^{\alpha}-1) + \alpha^2(\tau^{\alpha}+1)
        \\
        &\quad\quad
        +\tau(\tau^{\alpha}-1-\alpha(\tau^{\alpha}+1))
    )
    .
\end{align*}
Here, $\frac{\alpha\tau^{\alpha-2}}{(\tau^{\alpha}-1)^3}>0$.
Since $\alpha\geq1$, we have:
\begin{align*}
    &\alpha (\tau^{\alpha}-1) + \alpha^2(\tau^{\alpha}+1)+\tau(\tau^{\alpha}-1-\alpha(\tau^{\alpha}+1))
    \\
    &
    \geq
    1\times (\tau^{\alpha}-1) + 1\times\alpha(\tau^{\alpha}+1)+1(\tau^{\alpha}-1-\alpha(\tau^{\alpha}+1))
    \\
    &
    =2(\tau^{\alpha}-1) \geq 0
    .
\end{align*}
Namely, $\kappa(\tau)$ is convex.

\subsection{Proof of Proposition \ref{proposition: LogLogistic Example}}
\label{proposition: LogLogistic Example proof}

We drop the subscript $a$ for notational simplicity.
We start by computing $\kappa(T)$.
Recall the \CDF of the Log-Logistic distribution: $F_{C}(t) = \frac{1}{1+(t/\alpha)^{-\beta}}$.
Thus, $1-F_{C}(t) = \frac{1}{1+(t/\alpha)^{\beta}}$.
We compute $\expValDist{\min\{C, T\}}{C}$:
\begin{align}
    \nonumber
    \expValDist{\min\{C, T\}}{C}
    &=
    \int_0^{T} (1-F_{C}(t)) dt
    \\
    \nonumber
    &=
    \int_0^{T} \frac{dt}{1+(t/\alpha)^{\beta}}
    \\
    \label{eq: LL cond exp helper 1}
    &=
    \alpha\int_0^{T/\alpha} \frac{dz}{1+z^{\beta}}
    \\
    \label{eq: LL cond exp helper 2}
    &=
    \frac{\alpha}{\beta}\int_0^{ \frac{(T/\alpha)^\beta}{1+(T/\alpha)^\beta} } u^{\frac{1}{\beta}-1} (1-u)^{-\frac{1}{\beta}}du
    \\
    \nonumber
    &=
    \frac{\alpha}{\beta}\int_0^{ \frac{(T/\alpha)^\beta}{1+(T/\alpha)^\beta} } u^{\frac{1}{\beta}-1} (1-u)^{1-\frac{1}{\beta}-1}du
    \\
    \nonumber
    &=
    \frac{\alpha}{\beta}
    \incompleteBetaFunc{ \frac{(T/\alpha)^\beta}{1+(T/\alpha)^\beta} }{ \frac{1}{\beta} }{ 1-\frac{1}{\beta} }
    \\
    \nonumber
    &=
    \frac{\alpha}{\beta}
    \incompleteBetaFunc{F_C(T)}{\frac{1}{\beta}}{1-\frac{1}{\beta}}
    .
\end{align}
Note that Eq. \eqref{eq: LL cond exp helper 1} and \eqref{eq: LL cond exp helper 2} follow from substituting $z= t/\alpha$ and $u = z^{\beta}/(1+z^{\beta})$, respectively.
Thus, $\kappa(T) = (1+(T/\alpha)^{\beta})\frac{\alpha}{\beta} \incompleteBetaFunc{F_C(T)}{\frac{1}{\beta}}{1-\frac{1}{\beta}}$.

Recall that $f_C(T) = \frac{\beta/\alpha \times (T/\alpha)^{\beta-1}}{(1+(T/\alpha)^{\beta})^2}$.
For the second part, we derive $\kappa(T)$ with respect to $T$:
\begin{align*}
    \partialDerive{T}&\left(
        (1+(T/\alpha)^{\beta}) \frac{\alpha}{\beta} \incompleteBetaFunc{F_C(T)}{\frac{1}{\beta}}{1-\frac{1}{\beta}}
    \right)
    \\
    &=
    \frac{\beta}{\alpha} \left(\frac{T}{\alpha}\right)^{\beta-1} \frac{\alpha}{\beta} \incompleteBetaFunc{F_C(T)}{\frac{1}{\beta}}{1-\frac{1}{\beta}}
    \\
    &\quad
    +
    \left(1+\left(\frac{T}{\alpha}\right)^{\beta}\right) \frac{\alpha}{\beta} (F_C(T))^{\frac{1}{\beta}-1} (1-F_C(T))^{\frac{1}{\beta}} f_C(T)
    \\
    &=
    \left(\frac{T}{\alpha}\right)^{\beta-1} \incompleteBetaFunc{F_C(T)}{\frac{1}{\beta}}{1-\frac{1}{\beta}}
    \\
    &\quad
    +
    \left(\frac{T}{\alpha}\right)^{\beta-1} \frac{ (F_C(T))^{\frac{1}{\beta}-1} (1-F_C(T))^{\frac{1}{\beta}} }{1+\left(\frac{T}{\alpha}\right)^{\beta}}
    .
\end{align*}
Notably, this is the sum of two positive functions for any $T > 0$, so $\partialDerive{T}(\kappa(T)) \geq 0$ for any $T>0$.

For the third part, recall that $\expVal{C} = \alpha \incompleteBetaFunc{1}{1-\frac{1}{\beta}}{1+\frac{1}{\beta}}$ and substitute $T = \alpha$ in $\kappa(T)$.
Thus the condition $\kappa \leq \expVal{C}$ is translated to $\frac{2}{\beta} \incompleteBetaFunc{\frac{1}{2}}{\frac{1}{\beta}}{1-\frac{1}{\beta}}$ $\leq$ $\incompleteBetaFunc{1}{1-\frac{1}{\beta}}{1+\frac{1}{\beta}}$, which holds for any $\beta\in(1,2]$ (see Figure \ref{figure: beta diff}).

\begin{figure}[!htbp]
    \centering
    \includegraphics[scale=0.7]{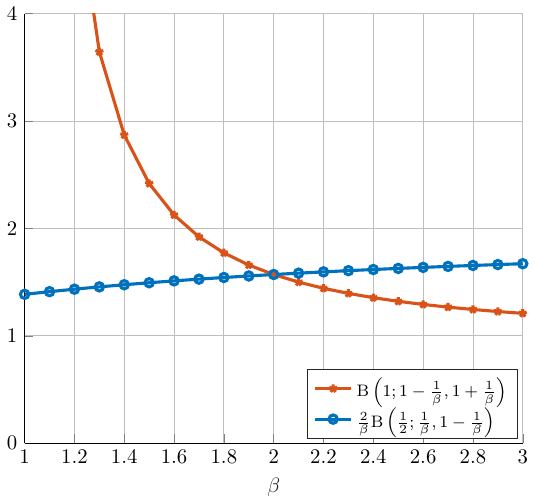}
    \vspace{-8pt}
    \caption{
        Comparing $\frac{2}{\beta} \incompleteBetaFunc{\frac{1}{2}}{\frac{1}{\beta}}{1-\frac{1}{\beta}}$ and $\incompleteBetaFunc{1}{1-\frac{1}{\beta}}{1+\frac{1}{\beta}}$, which reflect the relationship between $\kappa(\alpha)$ and $\expVal{C}$ for $\beta\in[1, 3]$ when $C\sim\mathrm{LogLogistic}(\alpha, \beta)$.
        For any $\beta\in[1, 2]$, $\kappa(\alpha) \leq \expVal{C}$.
        When $\beta > 2$, this inequality no longer holds.
    }
    \label{figure: beta diff}
\end{figure}

\fi

\bibliographystyle{IEEEtran}
\bibliography{references}

@ARTICLE{Nitinawarat2013_SHT_Argmax2_KLD_wProofs,
  author={Nitinawarat, Sirin and Atia, George K. and Veeravalli, Venugopal V.},
  journal={IEEE Transactions on Automatic Control}, 
  title={Controlled Sensing for Multihypothesis Testing}, 
  year={2013},
  volume={58},
  number={10},
  pages={2451-2464},
  doi={10.1109/TAC.2013.2261188}
}

@article{Naghshvar_Javidi2013_SHT_DynamicProgramming,
    author = {Mohammad Naghshvar and Tara Javidi},
    title = {{Active sequential hypothesis testing}},
    volume = {41},
    journal = {The Annals of Statistics},
    number = {6},
    publisher = {Institute of Mathematical Statistics},
    pages = {2703 -- 2738},
    year = {2013},
    doi = {10.1214/13-AOS1144},
    URL = {https://doi.org/10.1214/13-AOS1144}
}

@article{Naghshvar_Javidi2013_SHT_DynamicProgramming_Supp,
    author = {Mohammad Naghshvar and Tara Javidi},
    title = {{Supplement to “Active Sequential Hypothesis Testing”}},
    volume = {41},
    journal = {The Annals of Statistics},
    number = {6},
    publisher = {Institute of Mathematical Statistics},
    pages = {2703 -- 2738},
    year = {2013},
    doi = {10.1214/13-AOS1144SUPP},
    URL = {https://doi.org/10.1214/13-AOS1144SUPP}
}

@ARTICLE{Cohen_Zhao2015_SHT_AnomalyDetection,
  author={Cohen, Kobi and Zhao, Qing},
  journal={IEEE Transactions on Information Theory}, 
  title={Active Hypothesis Testing for Anomaly Detection}, 
  year={2015},
  volume={61},
  number={3},
  pages={1432-1450},
  doi={10.1109/TIT.2014.2387857}
}

@book{CoverThomas2006,
    author = {Cover, Thomas M. and Thomas, Joy A.},
    title = {Elements of Information Theory (Wiley Series in Telecommunications and Signal Processing)},
    year = {2006},
    isbn = {0471241954},
    publisher = {Wiley-Interscience},
    address = {USA}
}

@book{leongarcia2008,
  address = {Upper Saddle River, NJ},
  author = {Leon-Garcia, Alberto},
  edition = {Third},
  isbn = {9780131471221 0131471228},
  publisher = {Pearson/Prentice Hall},
  refid = {181079252},
  title = {Probability, Statistics, and Random Processes for Electrical Engineering},
  year = 2008
}

@article{Chernoff1959SequentialHT,
    author = {Herman Chernoff},
    title = {Sequential Design of Experiments},
    volume = {30},
    journal = {The Annals of Mathematical Statistics},
    number = {3},
    publisher = {Institute of Mathematical Statistics},
    pages = {755 -- 770},
    year = {1959},
    doi = {10.1214/aoms/1177706205},
    URL = {https://doi.org/10.1214/aoms/1177706205}
}

@book{siegmund2013sequential,
  title={Sequential Analysis: Tests and Confidence Intervals},
  author={Siegmund, D.},
  isbn={9781475718621},
  lccn={85007942},
  series={Springer Series in Statistics},
  url={https://books.google.co.il/books?id=bGTTBwAAQBAJ},
  year={2013},
  publisher={Springer New York}
}

@ARTICLE{Bai_Katewa_Gupta_Huang2015_Stochastic_Source_Selection,
  author={Bai, Cheng-Zong and Katewa, Vaibhav and Gupta, Vijay and Huang, Yih-Fang},
  journal={IEEE Transactions on Signal Processing}, 
  title={A Stochastic Sensor Selection Scheme for Sequential Hypothesis Testing With Multiple Sensors}, 
  year={2015},
  volume={63},
  number={14},
  pages={3687-3699},
  doi={10.1109/TSP.2015.2425804}
}

@article{Wald_1945_SHT,
 ISSN = {00034851},
 URL = {http://www.jstor.org/stable/2235829},
 author = {A. Wald},
 journal = {The Annals of Mathematical Statistics},
 number = {2},
 pages = {117--186},
 publisher = {Institute of Mathematical Statistics},
 title = {Sequential Tests of Statistical Hypotheses},
 urldate = {2024-06-13},
 volume = {16},
 year = {1945}
}

@inproceedings{Gan_Jia_Li2021_Decision_Tree_SHT,
 author = {Gan, Kyra and Jia, Su and Li, Andrew},
 booktitle = {Advances in Neural Information Processing Systems},
 editor = {M. Ranzato and A. Beygelzimer and Y. Dauphin and P.S. Liang and J. Wortman Vaughan},
 pages = {5012--5024},
 publisher = {Curran Associates, Inc.},
 title = {Greedy Approximation Algorithms for Active Sequential Hypothesis Testing},
 url = {https://proceedings.neurips.cc/paper_files/paper/2021/file/27e9661e033a73a6ad8cefcde965c54d-Paper.pdf},
 volume = {34},
 year = {2021}
}

@article{Armitage1950_SHT_MultipleHypotheses,
 ISSN = {00359246},
 URL = {http://www.jstor.org/stable/2983839},
 author = {P. Armitage},
 journal = {Journal of the Royal Statistical Society. Series B (Methodological)},
 number = {1},
 pages = {137--144},
 publisher = {[Royal Statistical Society, Wiley]},
 title = {Sequential Analysis with More than Two Alternative Hypotheses, and its Relation to Discriminant Function Analysis},
 urldate = {2024-08-21},
 volume = {12},
 year = {1950}
}

@INPROCEEDINGS{Citron_Cohen_Zhao2024_DGF_on_Hidden_Markov_Chains,
  author={Citron, Levli and Cohen, Kobi and Zhao, Qing},
  booktitle={{IEEE} International Symposium on Information Theory ({ISIT})}, 
  title={Anomaly Search of a Hidden Markov Model}, 
  year={2024},
  volume={},
  number={},
  pages={3684-3688},
  doi={10.1109/ISIT57864.2024.10619616}
}

@ARTICLE{Gurevich2019_EEST,
  author={Gurevich, Andrey and Cohen, Kobi and Zhao, Qing},
  journal={IEEE Transactions on Signal Processing}, 
  title={Sequential Anomaly Detection Under a Nonlinear System Cost}, 
  year={2019},
  volume={67},
  number={14},
  pages={3689-3703},
  doi={10.1109/TSP.2019.2918981}
}

@ARTICLE{Joseph_DeepLearining1,
  author={Joseph, Geethu and Zhong, Chen and Gursoy, M. Cenk and Velipasalar, Senem and Varshney, Pramod K.},
  journal={IEEE Sensors Journal}, 
  title={Anomaly Detection via Learning-Based Sequential Controlled Sensing}, 
  year={2024},
  volume={24},
  number={13},
  pages={21025-21037},
  doi={10.1109/JSEN.2024.3399456}
}

@ARTICLE{Szostak2024_DeepLearining2,
  author={Szostak, Hadar and Cohen, Kobi},
  journal={IEEE Access}, 
  title={Deep Multi-Agent Reinforcement Learning for Decentralized Active Hypothesis Testing}, 
  year={2024},
  volume={},
  number={},
  pages={1-1},
  doi={10.1109/ACCESS.2024.3430392}
}

@misc{stamatelis2024_DeepLearining3,
      title={Single- and Multi-Agent Private Active Sensing: A Deep Neuroevolution Approach}, 
      author={George Stamatelis and Angelos-Nikolaos Kanatas and Ioannis Asprogerakas and George C. Alexandropoulos},
      year={2024},
      eprint={2403.10112},
      archivePrefix={arXiv},
      primaryClass={cs.AI},
      url={https://arxiv.org/abs/2403.10112}, 
}

@ARTICLE{Gafni2023_CompositeHT,
  author={Gafni, Tomer and Wolff, Benjamin and Revach, Guy and Shlezinger, Nir and Cohen, Kobi},
  journal={IEEE Transactions on Signal Processing}, 
  title={Anomaly Search Over Discrete Composite Hypotheses in Hierarchical Statistical Models}, 
  year={2023},
  volume={71},
  number={},
  pages={202-217},
  doi={10.1109/TSP.2023.3242074}
}

@ARTICLE{Dragalin_etAl_1999_MSPRT_AsympOpt,
  author={Dragalin, V.P. and Tartakovsky, A.G. and Veeravalli, V.V.},
  journal={IEEE Transactions on Information Theory}, 
  title={Multihypothesis Sequential Probability Ratio Tests .{I}. Asymptotic Optimality}, 
  year={1999},
  volume={45},
  number={7},
  pages={2448-2461},
  doi={10.1109/18.796383}
}

@ARTICLE{Dragalin_etAl_2000_MSPRT_MeanSamplesApprox,
  author={Dragalin, V.P. and Tartakovsky, A.G. and Veeravalli, V.V.},
  journal={IEEE Transactions on Information Theory}, 
  title={Multihypothesis sequential probability ratio tests. {II}. Accurate asymptotic expansions for the expected sample size}, 
  year={2000},
  volume={46},
  number={4},
  pages={1366-1383},
  doi={10.1109/18.850677}
}

@INPROCEEDINGS{vershinin2025multistageactivesequentialhypothesis,
  author={Vershinin, George and Cohen, Asaf and Gurewitz, Omer},
  booktitle={IEEE International Symposium on Information Theory (ISIT)}, 
  title={Multi-Stage Active Sequential Hypothesis Testing with Clustered Hypotheses}, 
  year={2025},
  volume={},
  number={},
  pages={1-6},
  doi={10.1109/ISIT63088.2025.11195532}
}

@misc{vershinin2025iterativehypothesispruningdistributionbased,
      title={Iterative Hypothesis Pruning and Distribution-based Early Labeling for Sequential Hypothesis Testing}, 
      author={George Vershinin and Asaf Cohen and Omer Gurewitz},
      year={2025},
      eprint={2509.25908},
      archivePrefix={arXiv},
      primaryClass={cs.IT},
      url={https://arxiv.org/abs/2509.25908}, 
}

@article{lorden1970excess,
  title={On excess over the boundary},
  author={Lorden, Gary},
  journal={The Annals of Mathematical Statistics},
  volume={41},
  number={2},
  pages={520--527},
  year={1970},
  publisher={Institute of Mathematical Statistics}
}

@book{boyd_vandenberghe_2004,
    place={Cambridge},
    title={Convex Optimization},
    DOI={10.1017/CBO9780511804441},
    publisher={Cambridge University Press},
    author={Boyd, Stephen and Vandenberghe, Lieven},
    year={2004}
}

@INPROCEEDINGS{vershinin2025activesequentialhypothesistesting,
  author={Vershinin, George and Cohen, Asaf and Gurewitz, Omer},
  booktitle={IEEE International Conference on Acoustics, Speech and Signal Processing (ICASSP)}, 
  title={Active Sequential Hypothesis Testing With Non-Homogeneous Costs}, 
  year={2026},
  volume={},
  number={},
  pages={91-95},
  doi={10.1109/ICASSP55912.2026.11463839}
}

@article{Sheldon_etal_2005_reliability_IFR_DFR,
 ISSN = {00219002},
 URL = {http://www.jstor.org/stable/30040857},
 author = {Sheldon M. Ross and J. George Shanthikumar and Zegang Zhu},
 journal = {Journal of Applied Probability},
 number = {3},
 pages = {797--809},
 publisher = {Applied Probability Trust},
 title = {On Increasing-Failure-Rate Random Variables},
 urldate = {2025-12-14},
 volume = {42},
 year = {2005}
}

@ARTICLE{Crovella_Bestavros_1997_ParetoInternetDelay,
  author={Crovella, M.E. and Bestavros, A.},
  journal={IEEE/ACM Transactions on Networking}, 
  title={Self-similarity in World Wide Web traffic: evidence and possible causes}, 
  year={1997},
  volume={5},
  number={6},
  pages={835-846},
  doi={10.1109/90.650143}
}

@ARTICLE{GagoBenitez_et_al_2013_LogLogisticInternetDelay,
  author={Gago-Benítez, Ana and Fernández-Madrigal, Juan-Antonio and Cruz-Martín, Ana},
  journal={IEEE Sensors Journal}, 
  title={Log-Logistic Modeling of Sensory Flow Delays in Networked Telerobots}, 
  year={2013},
  volume={13},
  number={8},
  pages={2944-2953},
  doi={10.1109/JSEN.2013.2263381}
}

@book{kleinrock1974queueing,
  title={Queueing Systems: Theory},
  author={Kleinrock, L.},
  isbn={9780471491101},
  lccn={74009846},
  series={A Wiley-Interscience publication},
  url={https://books.google.co.il/books?id=rUbxAAAAMAAJ},
  year={1974},
  publisher={Wiley}
}

@article{Wald_Wolfowitz_1948_SPRT_Optimality,
 ISSN = {00034851},
 URL = {http://www.jstor.org/stable/2235638},
 author = {A. Wald and J. Wolfowitz},
 journal = {The Annals of Mathematical Statistics},
 number = {3},
 pages = {326--339},
 publisher = {Institute of Mathematical Statistics},
 title = {Optimum Character of the Sequential Probability Ratio Test},
 urldate = {2026-05-14},
 volume = {19},
 year = {1948}
}

@book{durrett2019probability,
  title={Probability: theory and examples},
  author={Durrett, Rick},
  volume={49},
  isbn={1108592635},
  year={2019},
  publisher={Cambridge University Press}
}

\end{document}